\documentclass[12pt,preprint]{aastex}
\shorttitle{Swift/XRT Lightcurves and Unified Origin of
X-Rays}\shortauthors{Liang et al.} \slugcomment{}
\begin{document}

\title{A Comprehensive Analysis of {\em Swift}/XRT Data:
\\ IV. Single Power-Law Decaying Lightcurves vs. Canonical Lightcurves and Implications for a Unified Origin of X-rays}
\author{En-Wei Liang\altaffilmark{1}, Hou-Jun L\"{u} \altaffilmark{1}, Shu-Jin Hou\altaffilmark{1}, Bin-Bin
  Zhang\altaffilmark{2}, and Bing Zhang\altaffilmark{2}
}\altaffiltext{1}{Department of Physics, Guangxi University, Nanning 530004, China; lew@gxu.edu.cn}
\altaffiltext{2}{Department of Physics and Astronomy, University of Nevada, Las Vegas, NV 89154;
zhang@physics.unlv.edu}

\begin{abstract}
By systematically analyzing the {\em Swift}/XRT lightcurves detected before
2009 July, we find 19 lightcurves that monotonously decay as a single power law
(SPL) with an index of $1\sim 1.7$ from tens (or hundreds) of seconds to $\sim
10^{5}$ seconds post the GRB trigger. They are apparently different from the
canonical lightcurves characterized by a shallow-to-normal decay transition. We
compare the observations of the prompt gamma-rays and the X-rays for these two
samples of GRBs (SPL vs. canonical). No statistical difference is found in the
prompt gamma-ray properties for the two samples. The X-ray properties of the
two samples are also similar, although the SPL sample tend to have a slightly
lower neutral hydrogen absorption column for the host galaxies and a slightly
larger energy release compared with the canonical sample. The SPL XRT
lightcurves in the burst frame gradually merge into a conflux, and their
luminosities at $10^5$ seconds are normally distributed at $\log L/{\rm ergs\
s}^{-1}=45.6\pm 0.5$. The normal decay segment of the canonical XRT lightcurves
has the same feature. Similar to the normal decay segment, the SPL lightcurves
satisfy the closure relations and therefore can be roughly explained with
external shock models. In the scenario that the X-rays are the afterglows of
the GRB fireball, our results indicate that the shallow decay would be due to
energy injection into the fireball and the total energy budget after injection
for both samples of GRBs is comparable. More intriguing, we find that a prior
X-ray emission model proposed by Yamazaki is more straightforward to interpret
the observed XRT data. We show that the zero times ($T_0$) of the X-rays are
$10^2\sim 10^5$ seconds prior to the GRB trigger for the canonical sample, and
satisfy a log-normal distribution. The negligible $T_0$'s of the SPL sample are
consistent with being the tail of $T_0$ distributions at low end, suggesting
that the SPL sample and the canonical sample may be from a same parent sample.
Referenced to $T_0$, the canonical XRT lightcurves well trace the SPL
lightcurves. The $T_0$'s of the canonical lightcurves in our analysis are
usually much larger than the offsets of the known precursors from the main
GRBs. If the prior emission hypothesis is real, the X-ray emission is better
interpreted within the external shock models based on the spectral and temporal
indices of the X-rays. The lack of detection of a jet-like break in most XRT
lightcurves implies that the opening angle of the prior emission jet would be
usually large.
\end{abstract}


\section{Introduction\label{sec:intro}}
{\em Swift}, a multi-wavelength gamma-ray burst (GRB) mission (Gehrels et al
2004) has led to great progress in understanding the nature of this phenomenon
(Zhang 2007). With the promptly slewing capability, the on-board X-ray
Telescope (XRT, Burrows et al. 2004) catches the very early X-rays following
the prompt gamma-rays. The X-ray observations led to the identification of a
canonical X-ray light curve, which is composed of four successive segments,
i.e. an initial steep decay segment (with a decay slope\footnote{Throughout the
paper the notation $f_\nu(t)\propto t^{-\alpha}\nu^{-\beta}$ is adopted}
$\alpha_1>2$), a shallow decay segment ($\alpha_2<0.75$), a normal decay
($\alpha_3\sim 1$) and a jet-like decay segment ($\alpha_4>1.5$) (Tagliaferri
et al. 2005; Zhang et al. 2006; Nousek et al. 2006; O'Brien et al. 2006). The
lightcurves are usually superimposed by erratic flares (Burrows et al. 2005;
Chincarini et al. 2007; Falcone et al. 2007), which may be produced by late
activities of the GRB central engine (Burrows et al. 2005; Zhang et al. 2006;
Fan \& Wei 2005; King et al. 2005; Dai et al. 2006; Perna et al. 2006; Proga \&
Zhang 2006). The initial steep decay segment is generally believed to be the
delayed photons from the high latitudes with respect to the line of sight upon
the abrupt cessation of emission in the prompt emission region (Fenimore et al.
1996; Kumar \& Panaitescu 2000; Dermer 2004; Dyks et al. 2005; Zhang et al.
2006; Liang et al. 2006; Yamazaki et al. 2006; Nousek et al. 2006; Panaitescu
et al. 2006a; Lazzati \& Begelman 2006; Zhang et al. 2007a, 2009a; Qin 2009;
but see Pe'er et al. 2006, Duran \& Kumar, 2009). The origin of the jet-like
steep-decay segment occasionally found in a few cases (Burrows \& Racusin 2006;
Liang et al. 2008; Racusin et al. 2009) is not fully understood. They may be
jet breaks, but the chromatic behavior observed in some GRBs posts the issue
regarding whether the observed X-ray and optical emissions share the same
origin (e.g. Panaitescu et al. 2006b; Liang et al. 2008).

The shallow decay (or plateau) segment is usually seen in the XRT lightcurves
(O'Brien et al. 2006a; Liang et al. 2007). The mechanism of this segment is
highly debated. Phenomenally, the canonical lightcurves can be well fitted with
a two-component model, i.e., a prompt emission component and an afterglow
component (Willingale et al. 2007; Ghisellini et al. 2008). The physical origin
of the specific function form proposed by Willingale et al. is, however, not
understood. The widely discussed model for the shallow decay component is
energy injection into the external forward shock either from an long lasting
central engine or from an ejecta with a wide distribution of Lorentz factors
(Zhang et al. 2006; Nousek et al. 2006; Panaitescu et al. 2006a). Such a
picture is supported by the fact that the spectral index does not change across
the break and that the segment after the break (normal decay segment) is
consistent with the closure relation of the forward shock model (Liang et al.
2007). Although some breaks are consistent with being achromatic, some others
show chromatic behavior across the break (Panaitescu et al. 2006b; Liang et al.
2007), suggesting that this model cannot interpret all the data.  Liang et al.
(2007) argued that the physical origin of the shallow decay segment may be
diverse and those plateaus that are followed by abrupt cutoffs might be of
internal origin (see also Troja et al. 2007). Ideas going far beyond the
traditional fireball models were also proposed. Ioka et al. (2006) argued that
there might be a weak prior emission before the GRB trigger, which modified the
medium density profile to produce the shallow decay phase. Shao \& Dai (2007)
interpreted the X-ray lightcurve as due to dust scattering of some prompt
X-rays. The model however predicts an evolution of the spectral indices which
is not observed in most afterglows
(Shen et al. 2008). The upscattering of the forward shock
photons by a trailing leptonic shell may also give an X-ray plateau (Panaitescu
2007), but the model is more suitable to interpret X-ray plateaus with sharp
cutoffs at the end. Uhm \& Beloborodov (2007) and Genet, Daigne \& Mochkovitch
(2007) suggested that both X-ray and optical afterglows are from a long-lived
reverse shock. Ghisellini et al. (2007) argued that the X-ray afterglows is
produced by late internal shocks, and the shallow-to-normal transition is due
to the jet effect in the prompt ejecta (see also Nava et al. 2007). Kumar et
al. (2008) proposed that the observed X-rays are directly related to the
accretion power from the central engine, and that the different power law
segments in the canonical lightcurves may be related to mass-accretion of
different layers of the progenitor stars. Most recently, Yamazaki (2009)
suggested that the X-ray emission is prior to the GRB trigger, which may be
powered by an earlier activity of the central engine. It may decay with a
single power law, but because the offset of the zero time point $T_0$, the
log-log lightcurve with the GRB trigger time as $T_0$ would display an
artificial shallow-to-normal decay transition. This model can interpret no
spectral change across the break, and also the chromatic behavior between X-ray
and optical, if the optical emission is powered by the ejecta launched during
the prompt emission. If correct, it would imply that the GRB activity may start
before the gamma-ray trigger.

It is interesting that the XRT lightcurves of a small fraction of {\em Swift} GRBs monotonously decay with a
single power-law (SPL), such as GRB 061007 (Schady et al. 2007). They are apparently different from the
majority of bursts that show the canonical lightcurves. This raises the issue about what factors make the
difference between the two groups of the XRT lightcurves. Are they really physically different or just due to
an uncovered artificial effect? We focus on this issue in this paper by systematically comparing the
properties of both the prompt gamma-ray and the X-ray emission properties of these two groups of GRBs. The
XRT data reduction, temporal and spectral anlysises, and sample selection are presented in \S2. We compare
the observations of both the prompt gamma-rays and the X-rays in the shallow-to-normal decay segment between
the two groups of GRBs in \S 3 and \S 4, respectively. We discuss possible implications from the results of
our analysis in \S 5, and present the conclusions in \S 6. A concordance cosmology with parameters $H_0 = 71$
km s$^{-1}$ Mpc$^{-1}$, $\Omega_M=0.30$, and $\Omega_{\Lambda}=0.70$ are adopted.

\section{DATA REDUCTION AND SAMPLE SELECTION\label{sec:data}}
The XRT data are downloaded from the {\em Swift} data archive. We developed a script to automatically
download and maintain all the XRT data. The HEAsoft packages, including Xspec, Xselect, Ximage, and {\em
Swift} data analysis tools, are used for the data reduction. We have developed an IDL code to automatically
process the XRT data for a given burst in any user-specified time interval. The details of our code have been
presented in Zhang et al. (2007a; Paper I) and Liang et al. (2007, 2008; Papers II and III). Our procedure is
briefly described as follows.

Our code first runs the XRT tool xrtpipeline to reproduce the XRT clean event data, and then makes pile-up
corrections with the same methods as discussed in Romano et al. (2006) (for the Window Timing [WT] mode data)
and Vaughan et al. (2006) (for the Photon Counting [PC] mode data). Both the source and background regions
are annuli (for PC) or rectangular annuli (for WT). The inner radius of the (rectangular) annuli are
dynamically determined by adjusting the inner radius of the annuli through fitting the source brightness
profiles with a King (1971) point spread function (for PC) or determined by the photon flux using the method
described in Romano et al 2006 (for WT). If the pipe-up effect is not significant, the source regions are in
the shape of a circle with radius $R$ = 20 pixels (for PC) or of a 40$\times$20 pixels rectangle (for WT)
centered at the bursts positions. The background regions have the same size as the source region, but has a
distance of 20 pixels away from the source regions. The exposure correction is also made with an exposure map
created by XRT tools xrtexpomap. By considering these corrections, the code extracts the
background-subtracted light curve and spectrum for the whole XRT data set. The signal-to-noise ratio for the
lightcurves is normally taken as $>3\sigma$, and it is flexibly adjusted depending on the source brightness.

\begin{figure*}
\includegraphics[angle=0,scale=0.50]{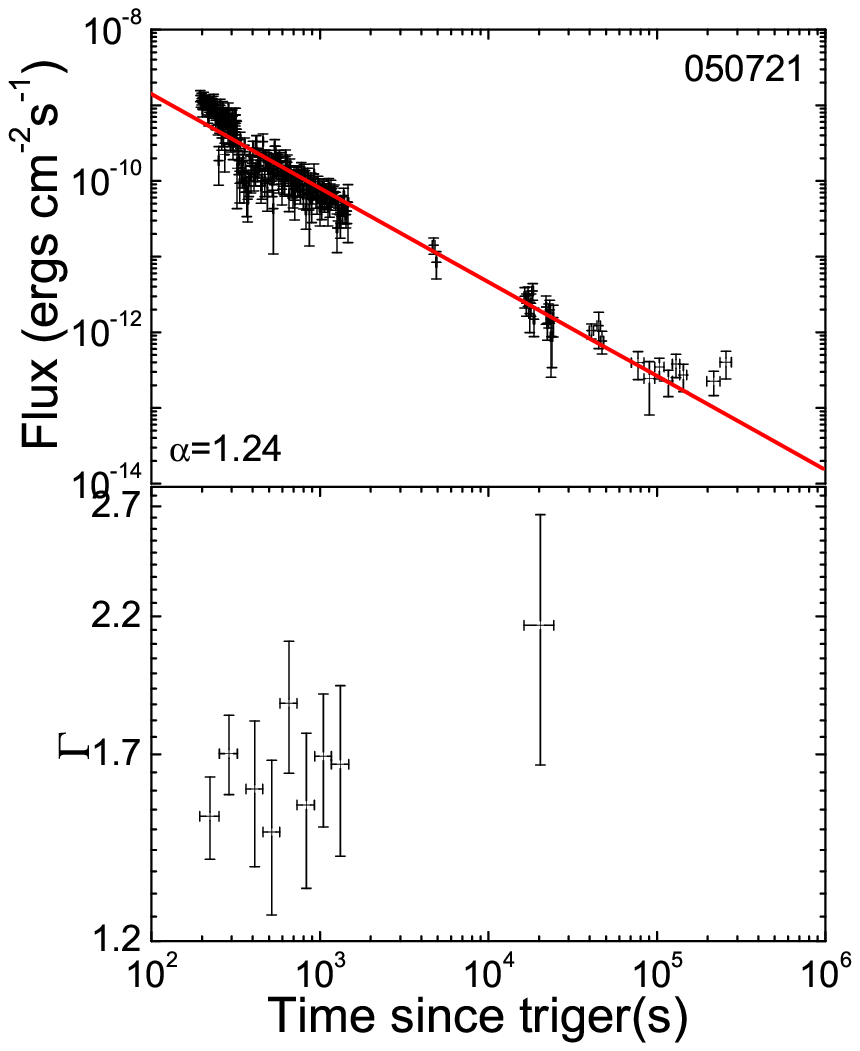}
\includegraphics[angle=0,scale=0.50]{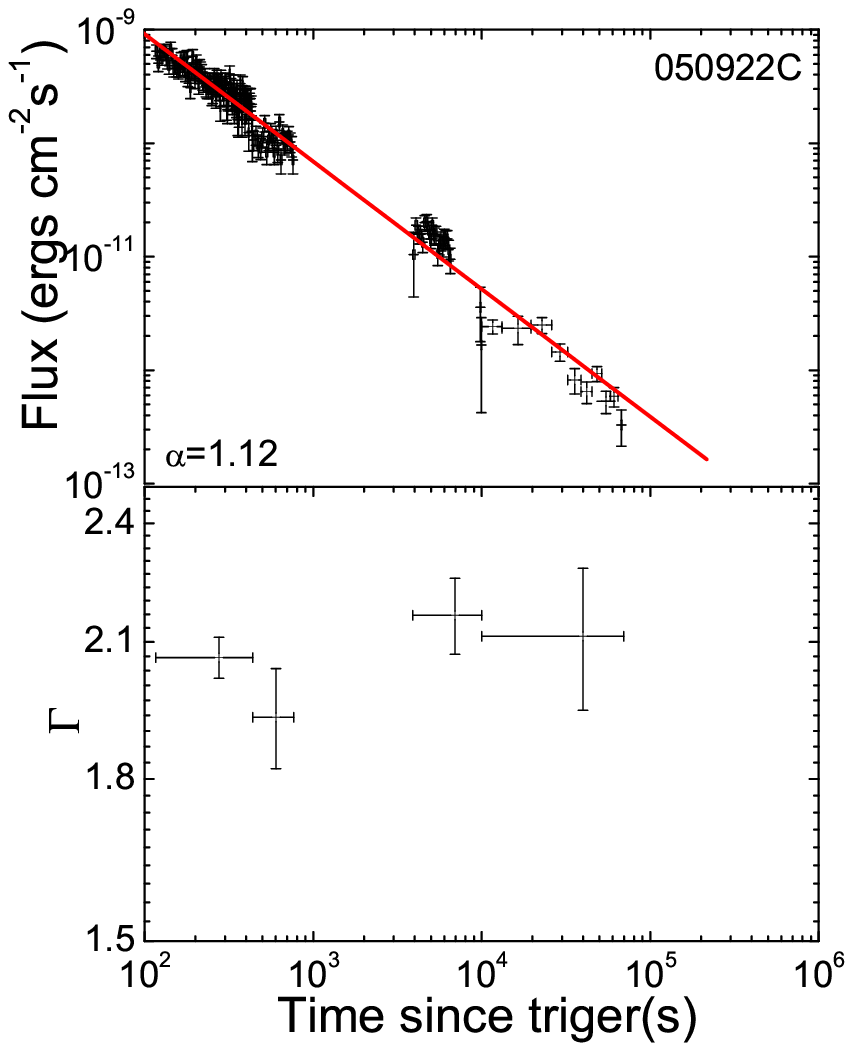}
\includegraphics[angle=0,scale=0.50]{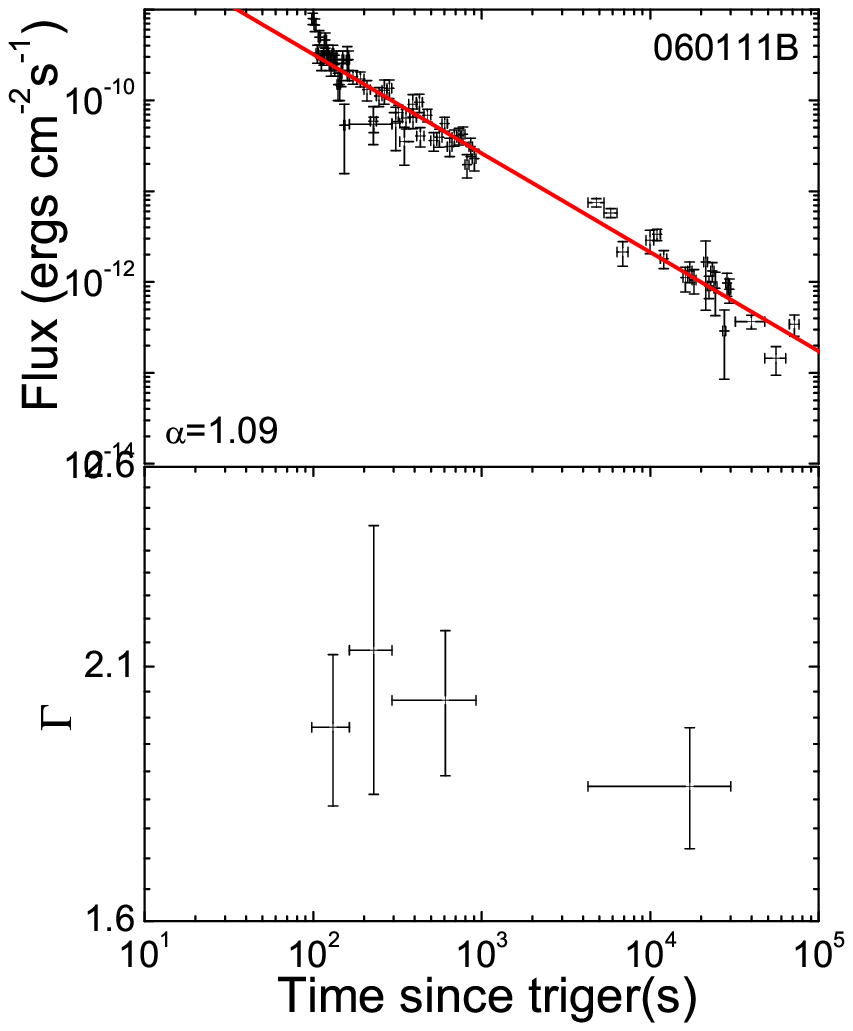}
\includegraphics[angle=0,scale=0.50]{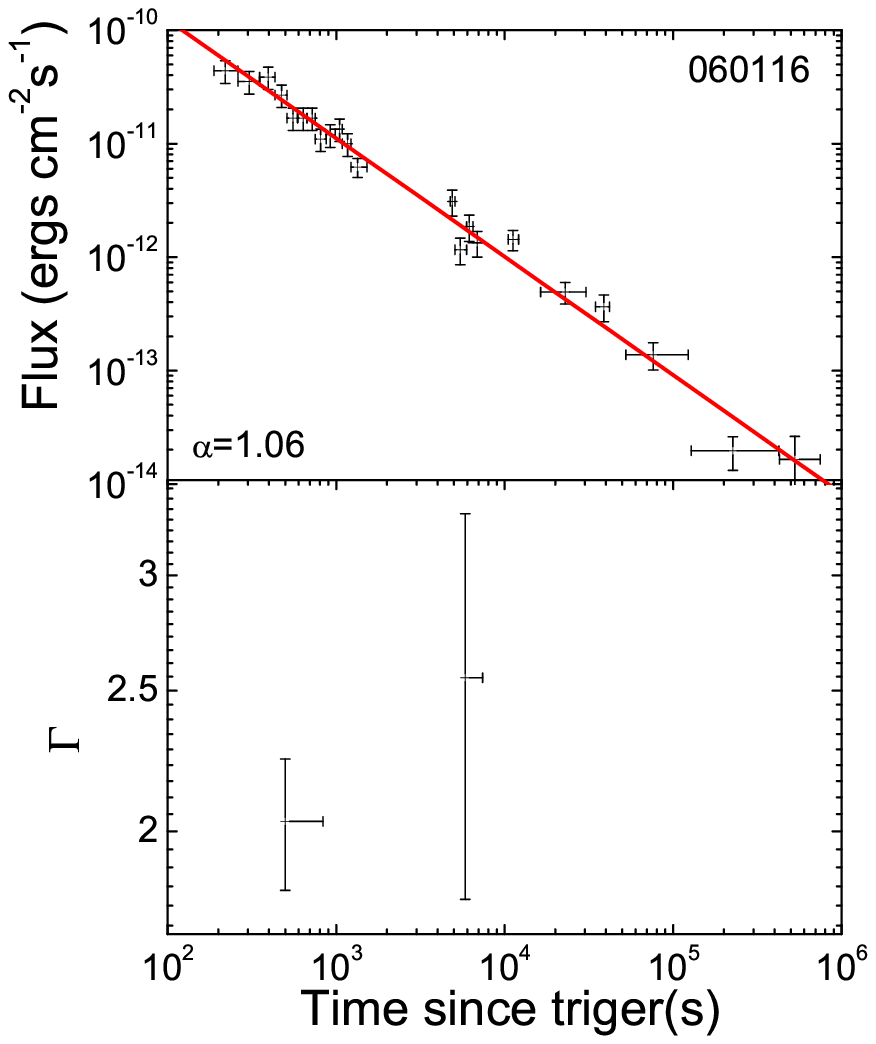}
\includegraphics[angle=0,scale=0.50]{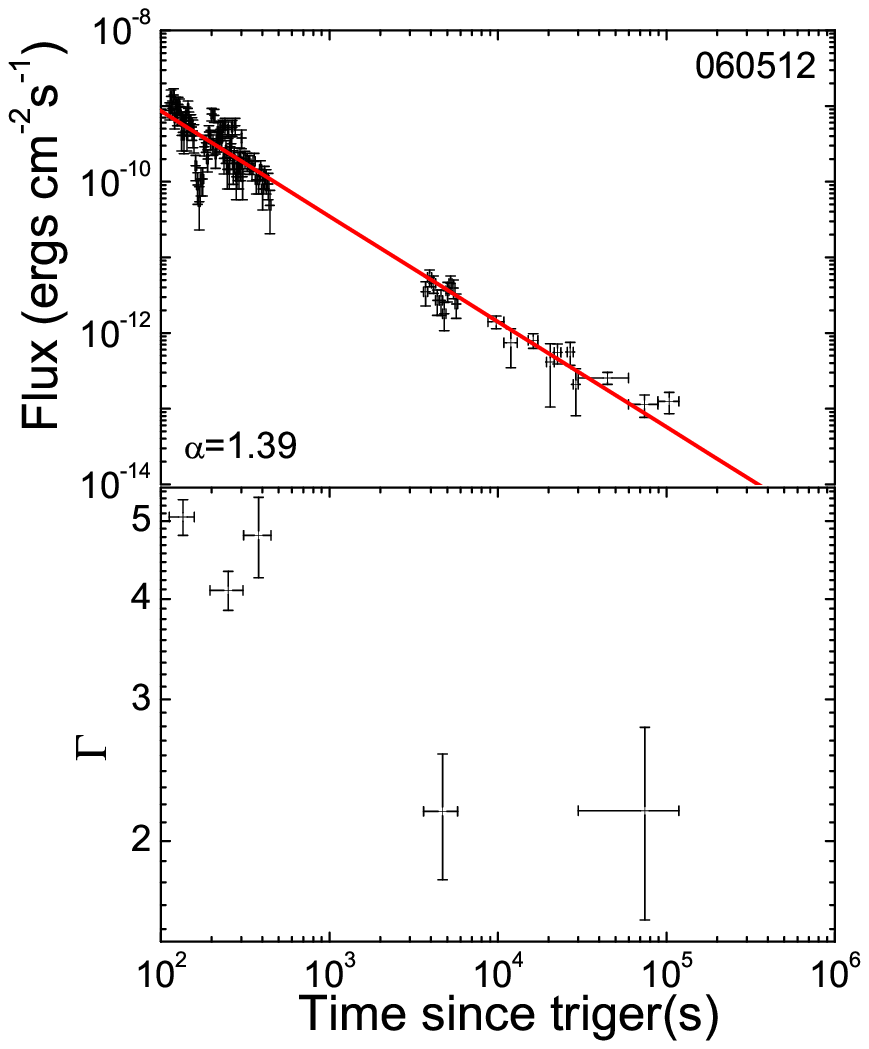}
\includegraphics[angle=0,scale=0.50]{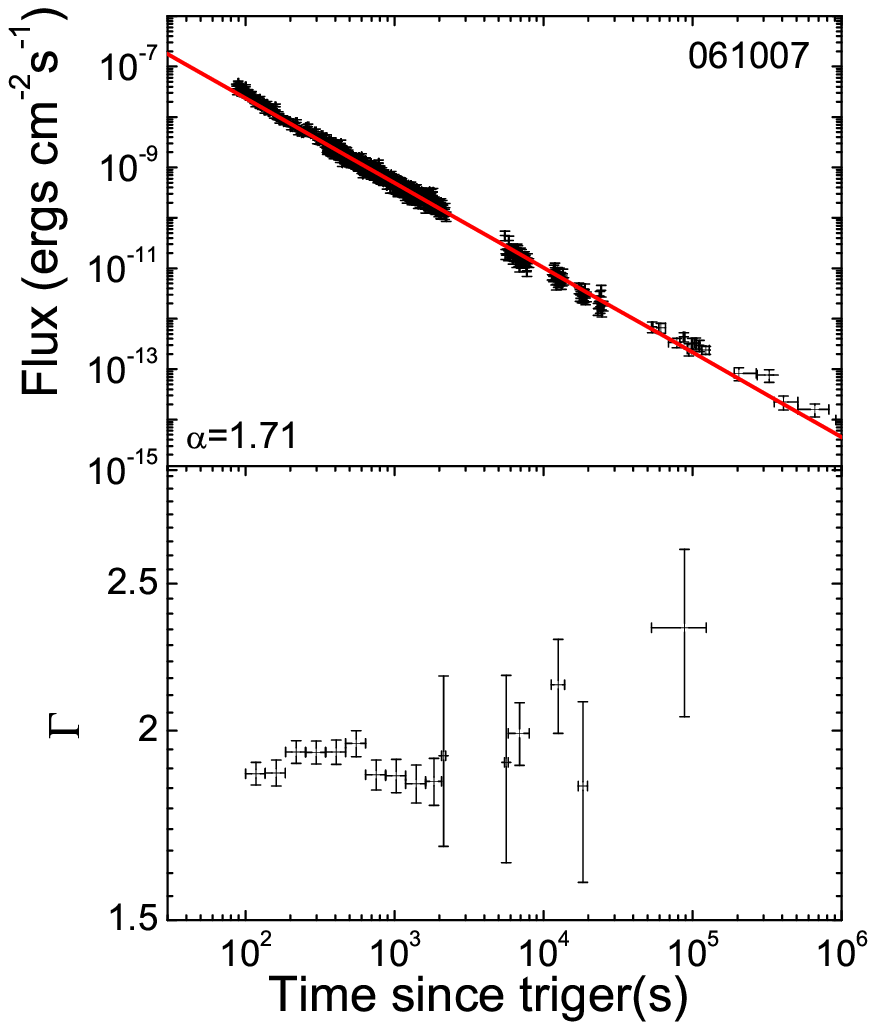}
\includegraphics[angle=0,scale=0.50]{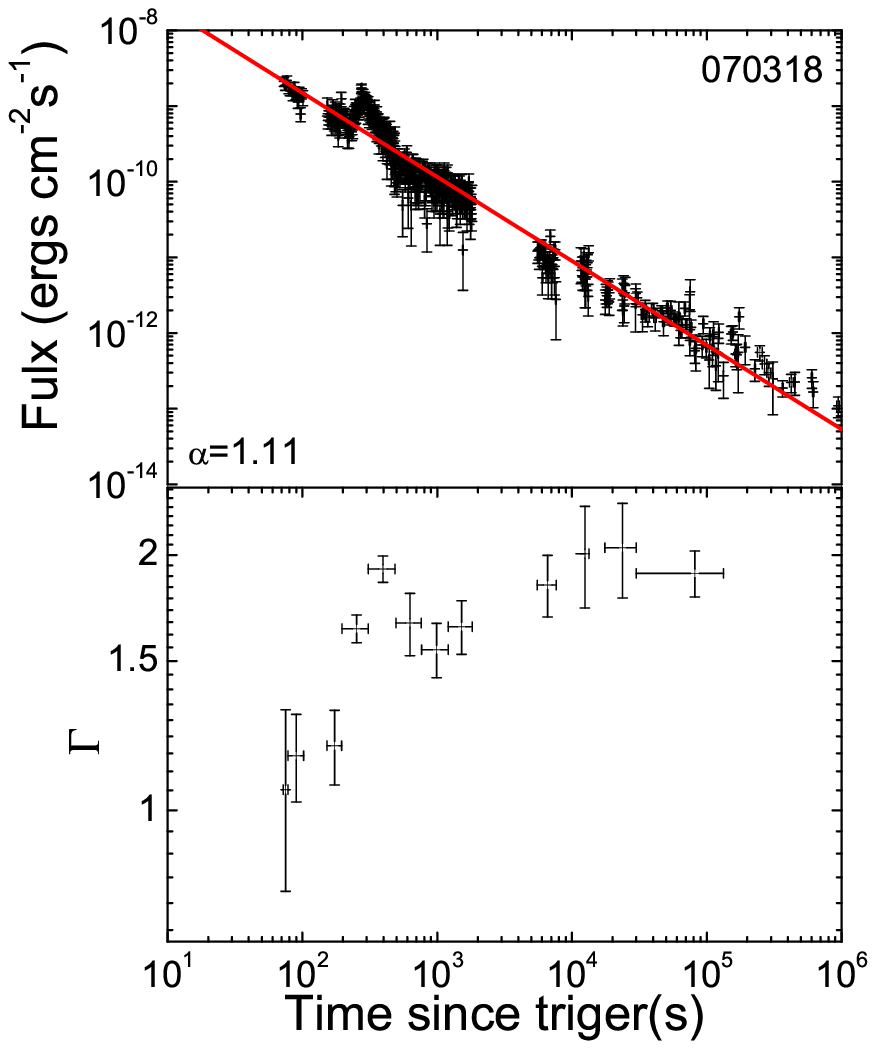}
\includegraphics[angle=0,scale=0.50]{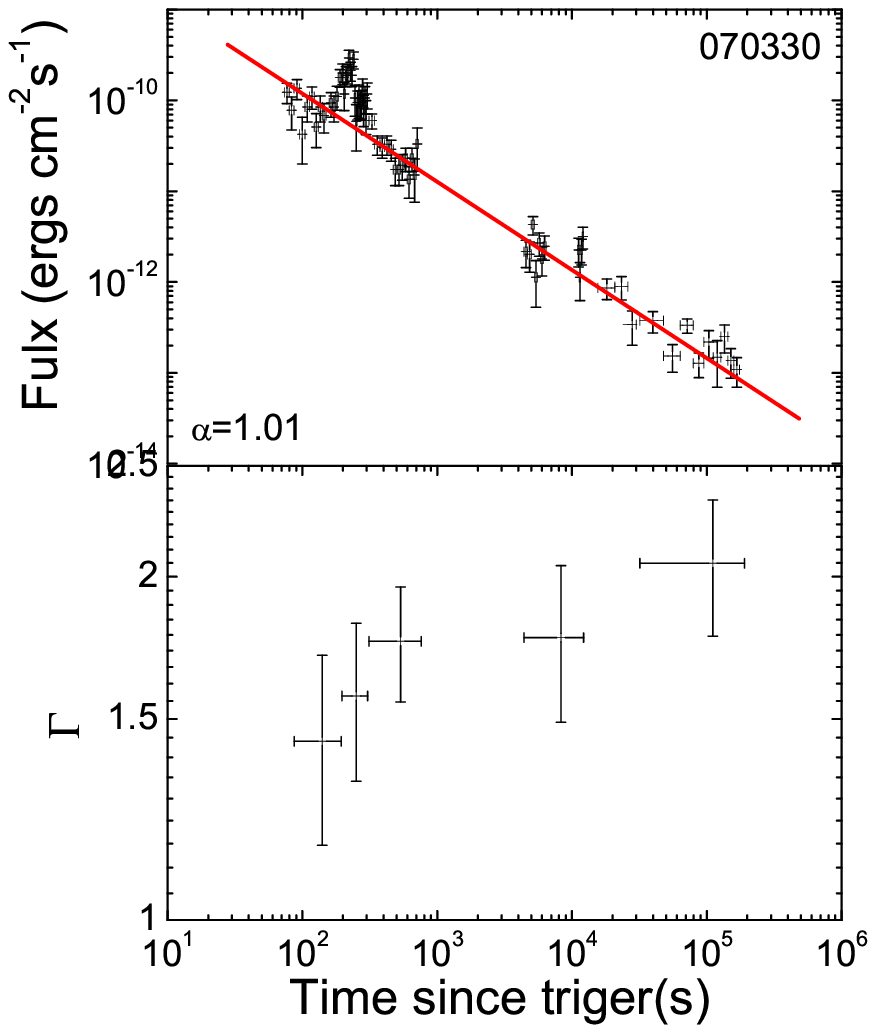}
\includegraphics[angle=0,scale=0.50]{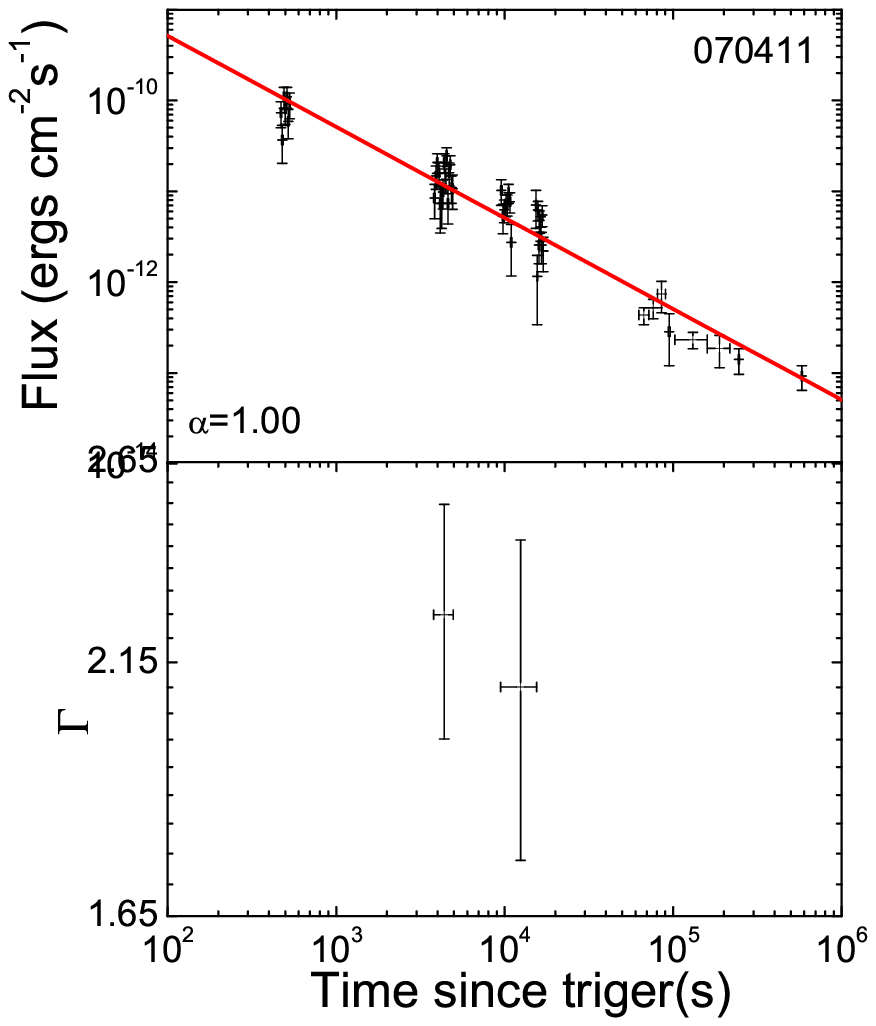}
\includegraphics[angle=0,scale=0.50]{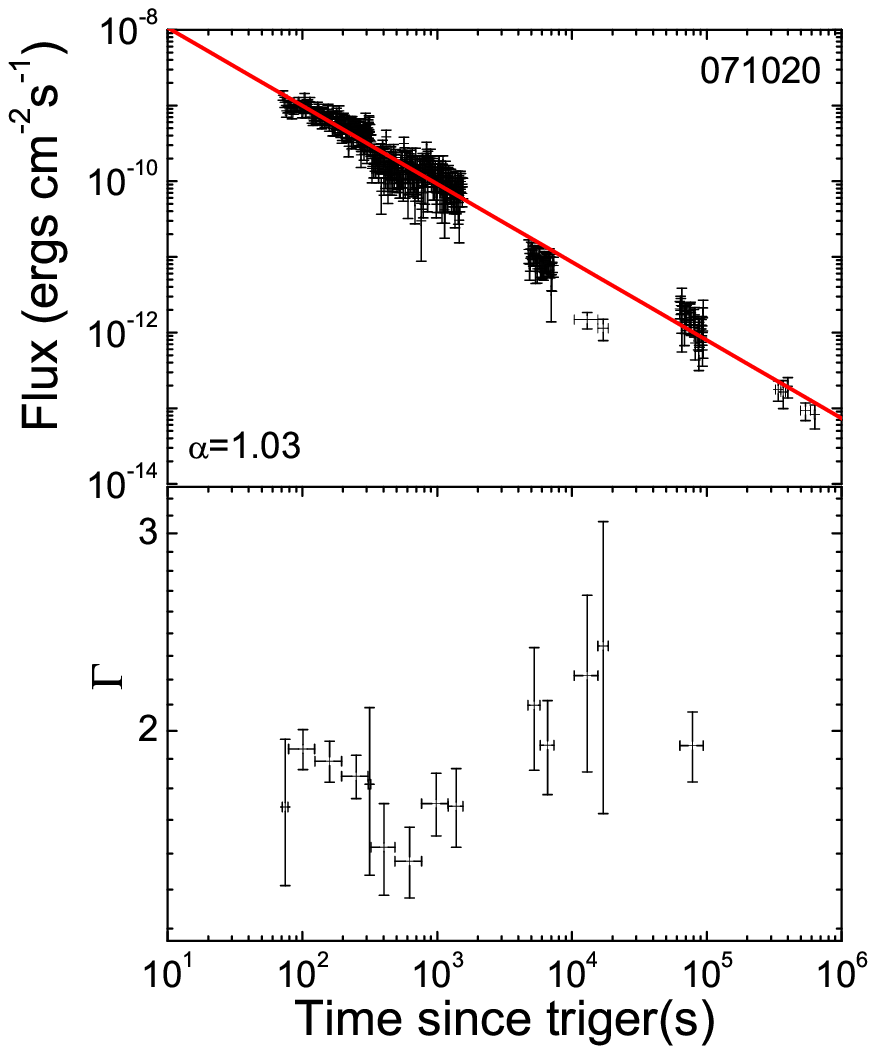}
\includegraphics[angle=0,scale=0.50]{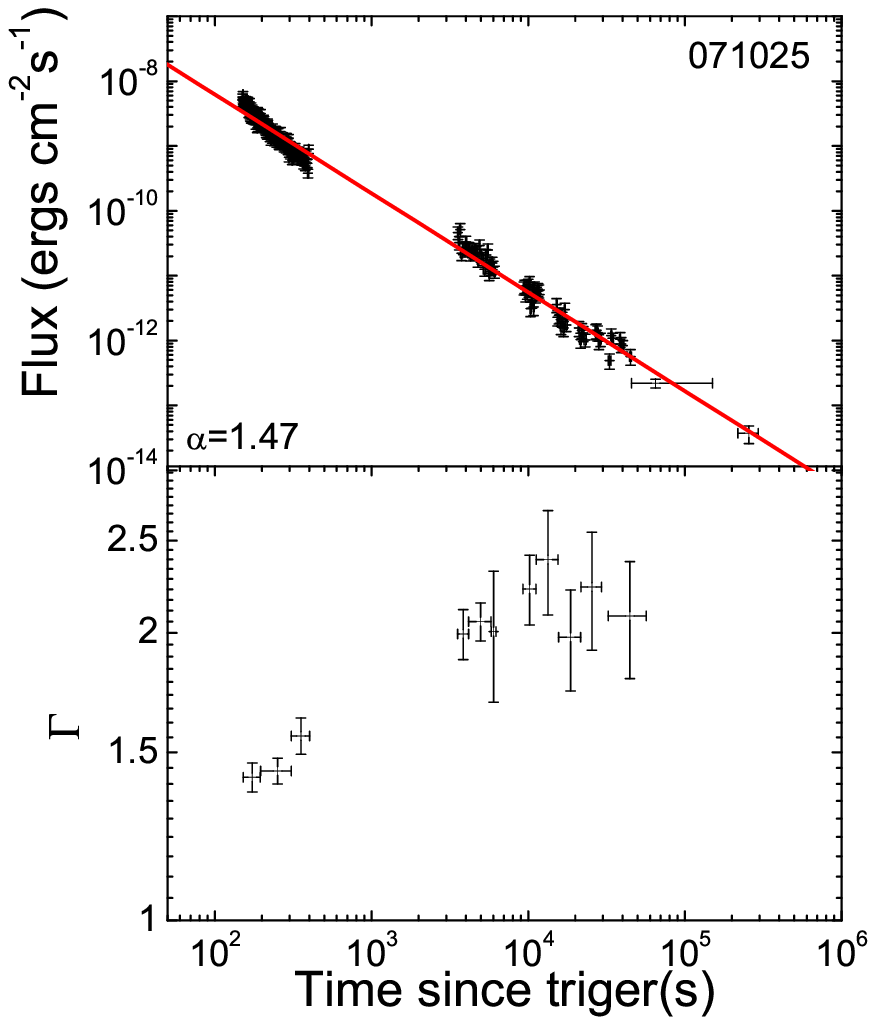}
\hfill
\includegraphics[angle=0,scale=0.50]{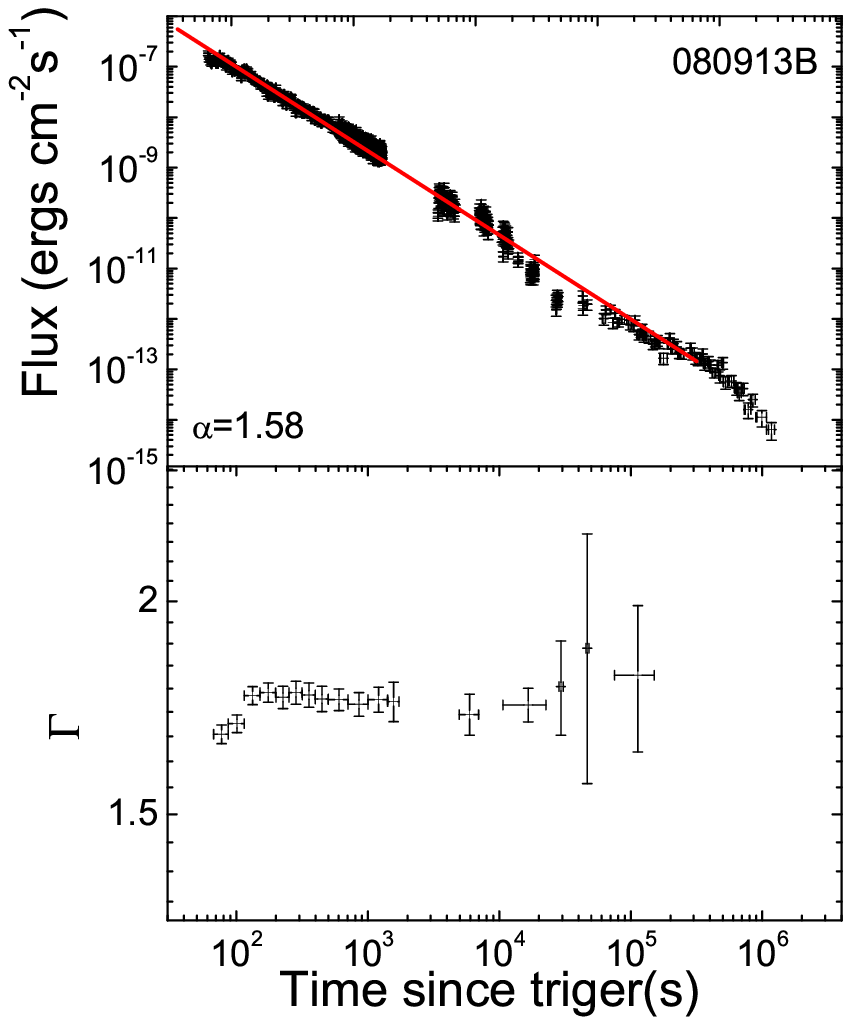}
\caption{XRT lightcurves (top panel) and spectral indices as a function of time
(bottom panel) of the 19 GRBs in the SPL sample. The solid lines are the best
fits to the SPL model.} \label{XRT_LC}
\end{figure*}

The XRT data observed from January 2005 to July 2009 for $\sim 400$ GRBs are
reduced with our code. We consider only long duration GRBs (or Type II GRBs;
Zhang 2006; Zhang et al. 2007b, 2009b). By visually going through the XRT
lightcurves for these bursts, we select those lightcurves that monotonously
decay with a SPL from tens or one hundred seconds to $\sim 10^5$ seconds post
the GRB triggers without a significant shallow-to-normal segment transition.
Only 19 solid cases are identified in our sample. Twelve out of the 19 GRBs
have redshift measurements\footnote{The redshift of GRB 060512 is uncertain. It
is reported as 0.4428 by Bloom et al (2006b) but as 2.1 by Starling et al.
(2006). We do not include this GRB in the redshift-known sample.}. By removing
the superimposed flares from the lightcurves, we fit the lightcurves with a
SPL. The lightcurves with our fits are shown in Fig. \ref{XRT_LC}, and the
power law indices are reported in Table 1. We perform a time-resolved spectral
analysis for the data with an absorbed power-law model, i.e.,
abs$\times$zabs$\times$power-law, where abs and zabs are the absorbtion models
for the Milky Way Galaxy and the GRB host galaxy, respectively, if the redshift
of the GRB is available. For a given GRB, we do not consider the evolution of
host galaxy $N_H^{\rm host}$, and fix it as the value derived from the fit to
the time-integrated spectrum. The values of $\Gamma_X$ and $N_H^{\rm host}$
derived from the fits to the time-integrated spectrum are reported in Table 1.
The time interval taken for the spectral fitting is dynamically determined by
the photons accumulated in this time interval, which is required to obtain a
photon index with more than $3\sigma$ significance. The time-resolved spectral
analysis for each burst is shown in Fig \ref{XRT_LC}. The BAT observations of
these bursts are collected from the published papers or GCN reports, and they
are summarized also in Table 1.

In order to compare the properties of these GRBs to the GRBs having a canonical XRT lightcurve, we compile a sample of
80 XRT lightcurves that have a well-sampled shallow-to-normal transition feature. The initial steep decay segment is
removed from these lightcurves since this segment is generally believed to be the GRB tail emission due to the curvature
effect as mentioned above. The redshifts of these bursts are also required in order to derive the properties of these
bursts in the burst frame. We do not include GRBs 060522, 060607A, and 070110 in our sample since they abruptly transit
to a very steep decay phase and they might have a different physical origin (Liang et al. 2007; Troja et al. 2007;
Kumar et al. 2008). GRB 060614 is also not included since it may belong to the category of compact star mergers
(Type I GRBs) (Gehrels et al. 2006; Zhang 2006; Zhang et al. 2007b, 2009b). We thus get 44 GRBs with redshift
measurement in out canonical sample. The observations for these bursts are summarized in Table 2.

\section{THE CHARACTERISTICS OF THE PROMPT GAMMA-RAYS}
\begin{figure*}
\includegraphics[angle=0,scale=0.80]{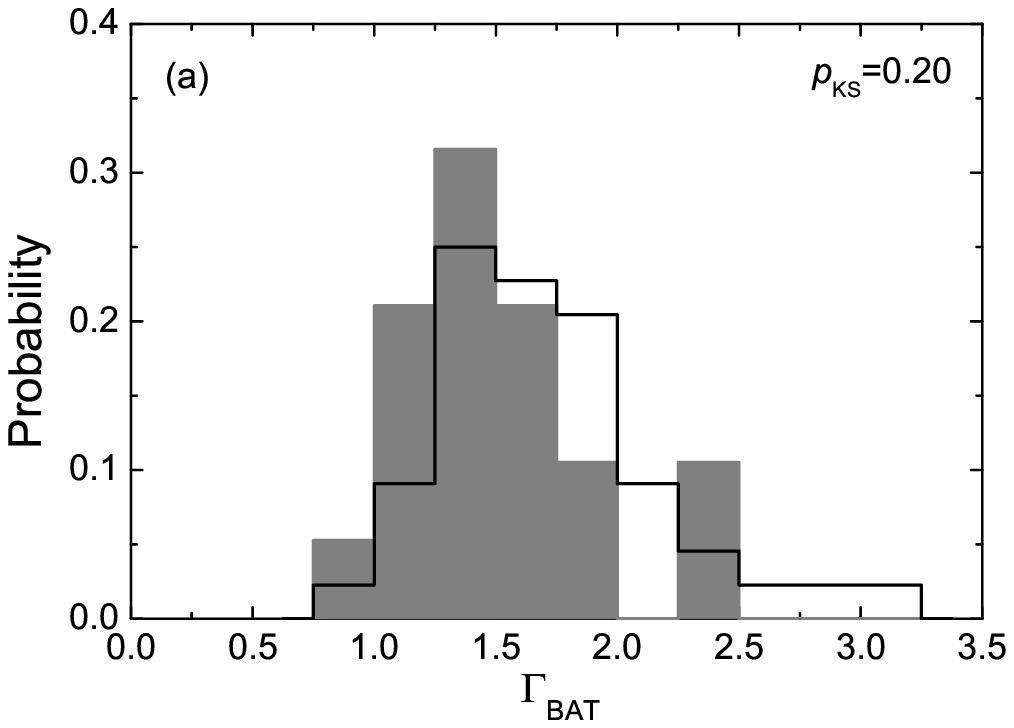}
\includegraphics[angle=0,scale=0.80]{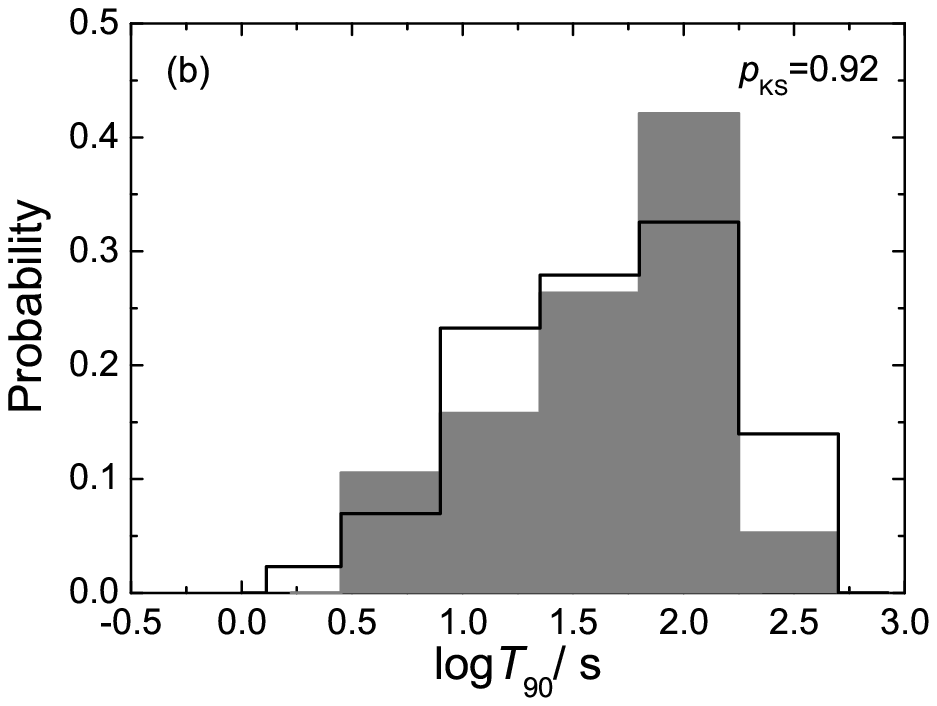}
\includegraphics[angle=0,scale=0.80]{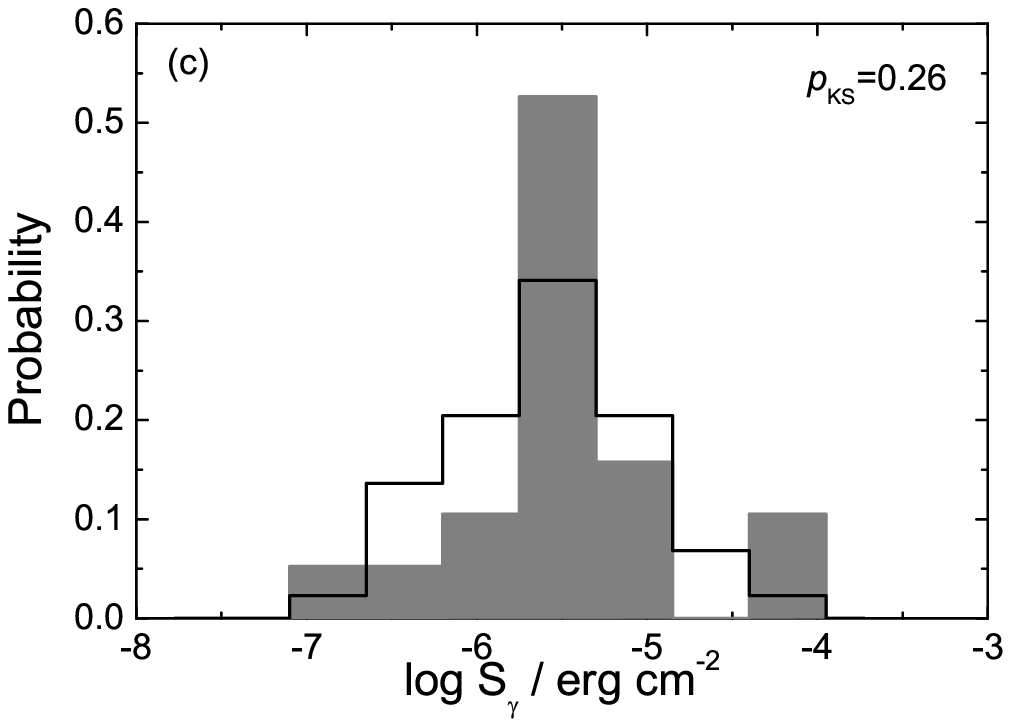}
\hfill
\includegraphics[angle=0,scale=0.80]{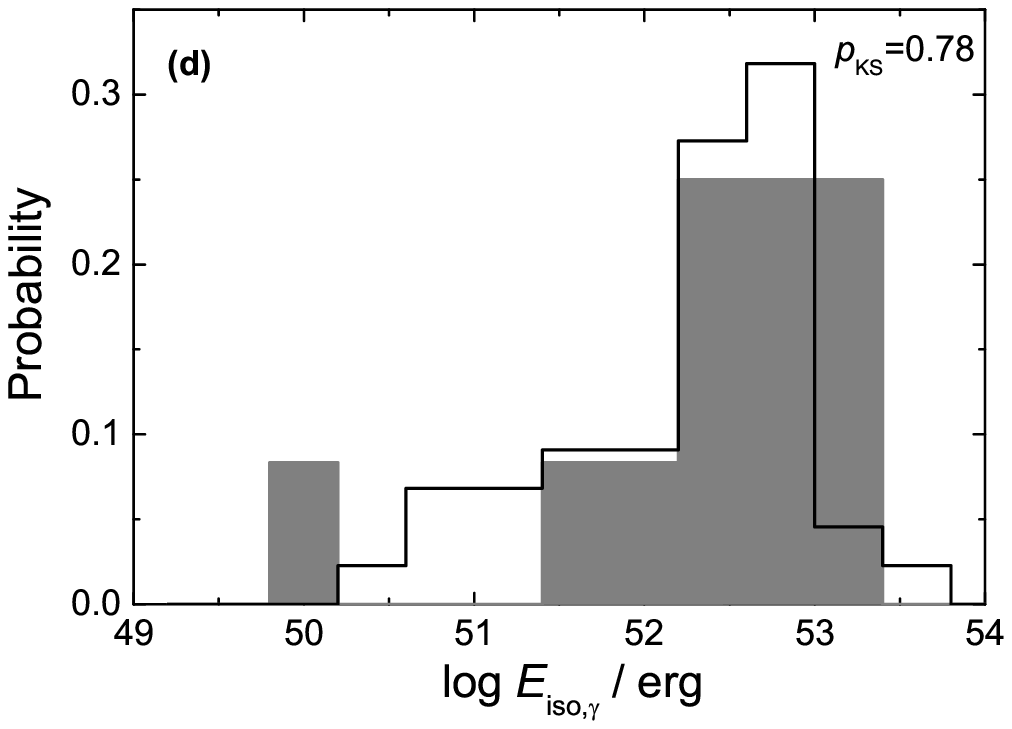}
\caption{Comparisons of the distributions of $\Gamma_{BAT}$, $T_{90}$,
$S_\gamma$, and $E_{iso,\gamma}$ between the SPL sample ({\em grey columns})
and the canonical sample ({\em solid line}). The K-S probability of consistency
for each comparison ($p_{KS}$) is also marked. } \label{Prompt_Dis}
\end{figure*}

The distributions of the photon index ($\Gamma_{\rm BAT}$), burst duration
($T_{90}$), gamma-ray fluence ($S_{\gamma}$), and isotropic gamma-ray energy
($E_{{\rm iso},\gamma}$) of the prompt gamma-rays in the BAT band for both the
SPL sample (solid) and the canonical sample (dashed) are shown in Fig.
\ref{Prompt_Dis}. We measure the difference of any pair of distributions with
the probability ($p_{KS}$) of the Kolmogorov-Smirnov test (K-S test). The
hypothesis that the two distributions are from the same parent sample is statistically
rejected if $p_{\rm KS}<10^{-4}$, and it is marginally rejected if
$10^{-4}<p_{\rm KS}<0.1$. The $p_{\rm KS}=1$ indicates that the two samples are
identical. The values of $p_{\rm KS}$ are marked in each panel of Fig.
\ref{Prompt_Dis}. It is found that the derived $p_{KS}$ are all greater than
0.1, indicating that there are no statistical differences of these
distributions between the two groups of GRBs.

\section{THE CHARACTERISTICS OF THE X-RAYS}
\begin{figure*}
\includegraphics[angle=0,scale=0.80]{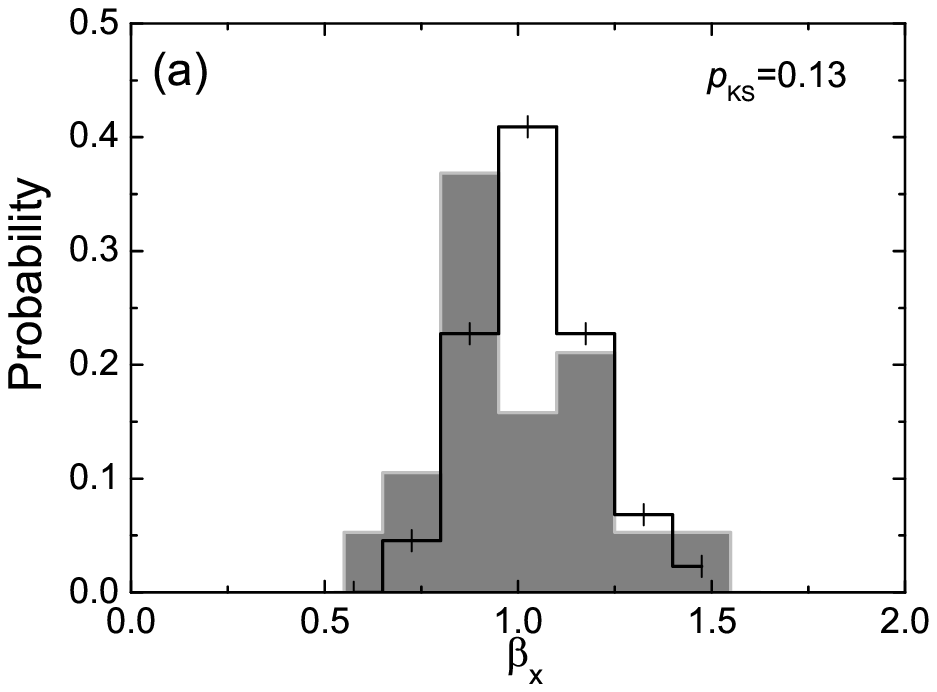}
\includegraphics[angle=0,scale=0.80]{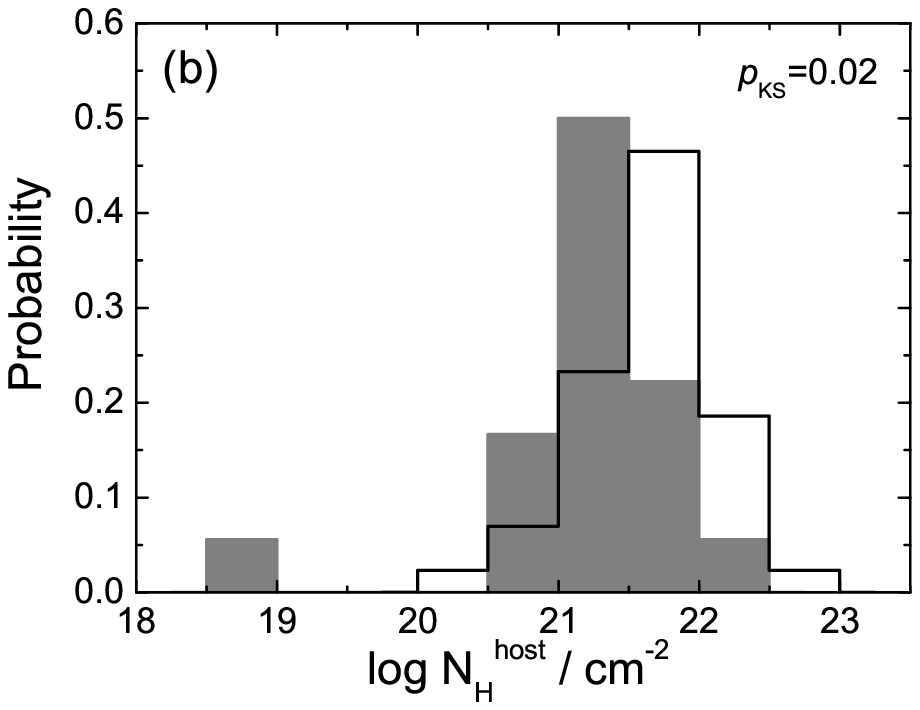}
\includegraphics[angle=0,scale=0.80]{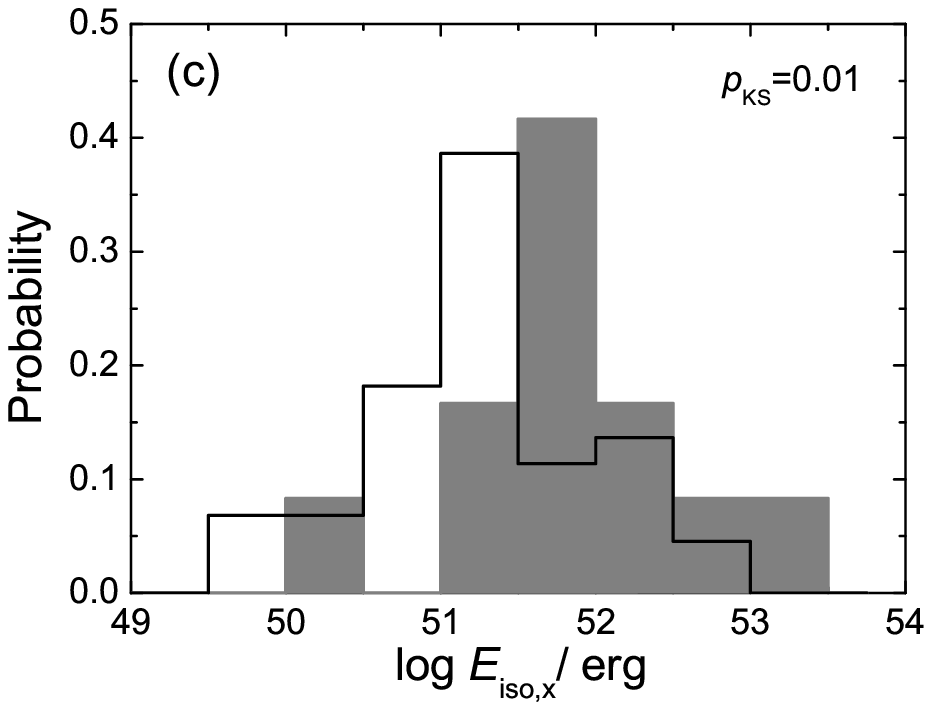}
\hfill
\includegraphics[angle=0,scale=0.80]{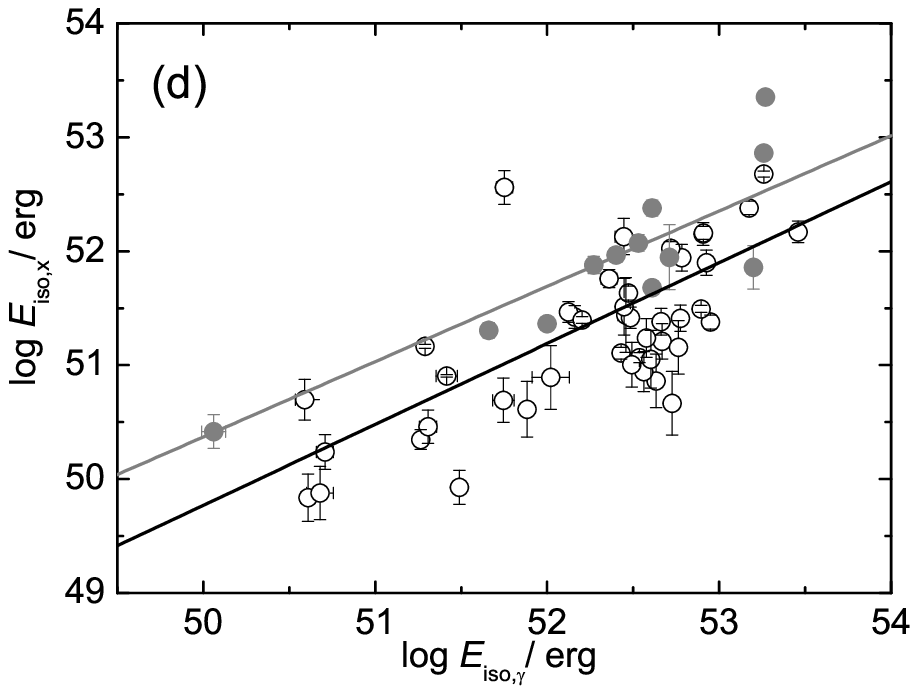}
\caption{Comparisons of the distributions of $\beta_X$, $N_H$ and $E_{\rm iso,
\ X}$ and the correlations between $E_{\rm iso,\ X}$ and $E_{\rm iso,\ \gamma}$
for the SPL sample ({\em grey columns or dots} ) and the canonical sample ({\em
solid lines or opened dots})} \label{Xray_Dis}
\end{figure*}

The distributions of the X-ray spectral index ($\beta_X$), $N_H^{\rm host}$,
and isotropic X-ray energy ($E_{\rm iso, X}$) in the XRT band  for the two
groups of GRBs are shown in Fig. \ref{Xray_Dis}. The comparison of the
correlation between $E_{iso,\gamma}$ and $E_{\rm iso, X}$ is also displayed in
Fig. \ref{Xray_Dis}. The $E_{\rm iso, X}$ is integrated from $T_{90}$ to $10^5$
seconds post the GRB trigger. The $E_{{\rm iso}, X}$ of the GRBs with a
canonical XRT lightcurve is calculated in the same time interval by
extrapolating the shallow decay segment to the time of $T_{90}$ without taking
the steep decay segment into account. The values of $p_{\rm KS}$ are also
marked in Fig. \ref{Xray_Dis}. The $p_{KS}$ for the $\beta_X$ distributions is
0.13, indicating no statistical difference of $\beta_X$ between the two groups
of GRBs. The distributions of both ${N_H}^{\rm host}$ and $E_{\rm iso, X}$ show
slight differences between the two samples, with $p_{KS}\sim 10^{-2}$. The
$N_H^{\rm host}$ of the SPL sample tend to have a lower $N_H$ and larger
$E_{\rm iso, X}$ than the canonical sample, but the slopes of the $E_{\rm iso,
X}-E_{\rm iso, \gamma}$ relations are almost the same for the two samples, as
displayed in  Fig. \ref{Xray_Dis}(d).

We derive the lightcurves in the burst frame for the two groups of GRBs, which
are presented with isotropic X-ray luminosity ($L$) as a function of $t/(1+z)$.
They are displayed in Figs. \ref{T0_LC}(a) and \ref{T0_LC}(b). It is found that
the SPL decay lightcurves merge into a conflux at around one day post the GRB
trigger. The distribution of the luminosities at $10^5$ seconds is shown in
Fig. \ref{Luminsoity_Dis}(a), with an average of $\log (L_{10^5{\rm s}}/{\rm
ergs\
 s}^{-2})=45.56\pm 0.55$. Interestingly, the late X-ray luminosity of the canonical sample also shares
the similar feature, with an average of $\log (L_{10^5{\rm s}}/{\rm ergs\
s}^{-2})=45.21\pm 0.56$. The K-S test for the comparison of the two samples
gives $p_{KS}=0.10$, indicating that there is no statistical difference between
the two distributions.

\section{Discussion}
As shown above, no statistical difference of the prompt gamma-rays between the two groups of GRBs is found, and their
spectral characteristics of the X-rays are also consistent with each other. These results likely suggest that the
X-rays observed in the two groups may have the same physical origin, and the apparent shallow decay segment in XRT
lightcurves would be due to extrinsic effects. We discuss possible explanations of the X-rays in this section. Although
some ideas that go far beyond the traditional fireball models were proposed to explain the shallow-to-normal decay
segment(see \S 1), the most popular model is the traditional long lasting energy injection scenario. On the other hand,
motivated by our analysis above, we suspect that the smooth shallow-to-normal segment may be due to the real starting
time of this emission component is prior to the GRB trigger time, as suggested by Yamazaki (2009). We will discuss
these two scenarios.

\subsection{The X-rays as the afterglow component of the prompt GRBs}\label{Sec:external}
\begin{figure*}
\includegraphics[angle=0,scale=0.8]{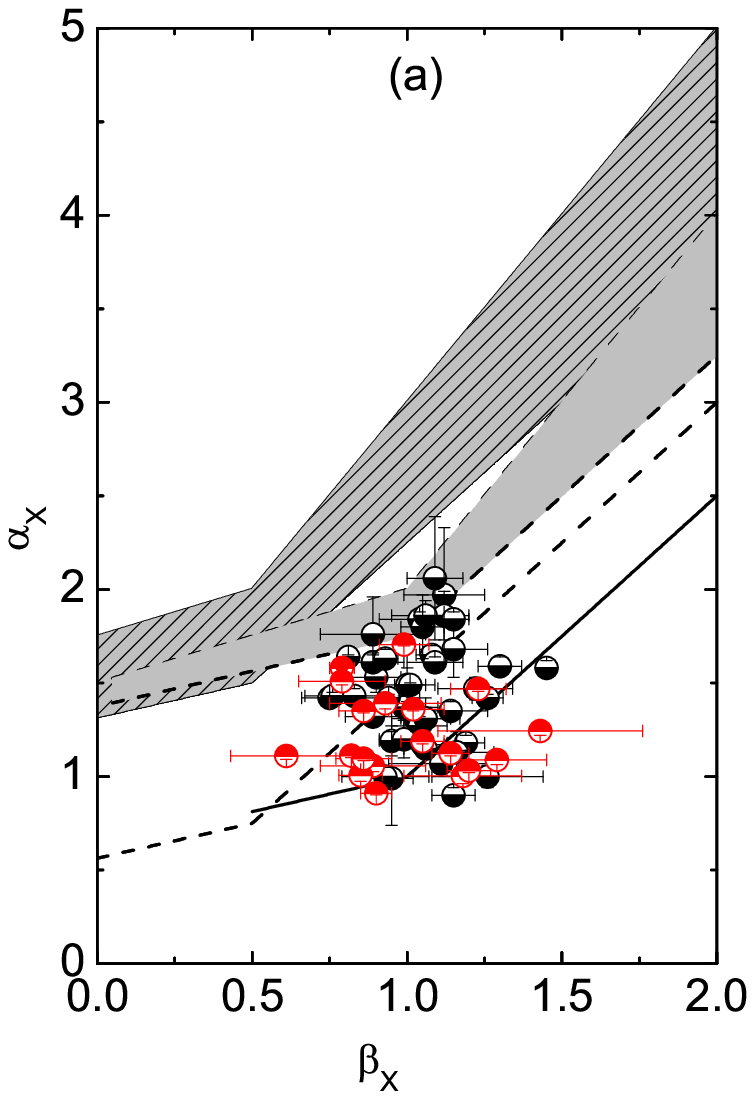}
\includegraphics[angle=0,scale=0.8]{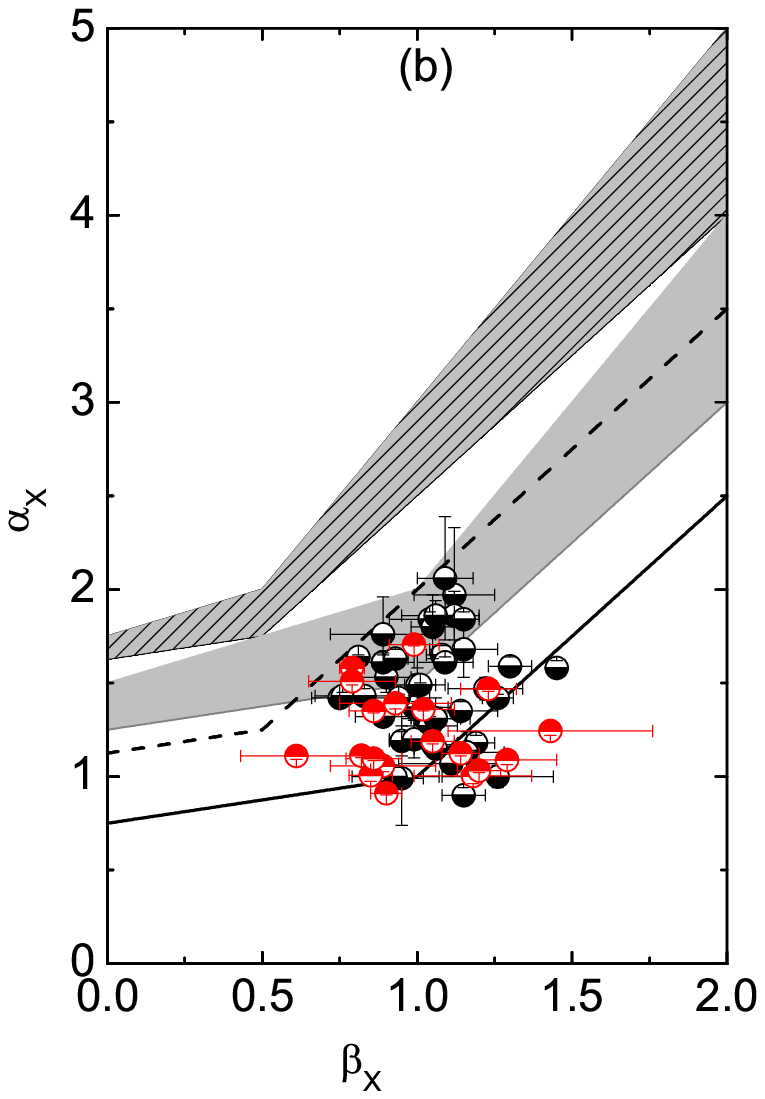}
 \caption{Temporal decay index against spectral index for the
SPL ({\em solid dots}) and the $T_0$-shifted canonical XRT lightcurves ({\em
open circles}). For comparison the closure relations of the external forward
shock afterglow models are overplotted. The solid lines and the shaded regions
are for the spectral regime I ($\nu_{x} > \rm{max}(\nu_m,\nu_c)$), and the
dashed lines and the hatch-shaded regions are for the spectral regime II
($\nu_m<\nu_{x}<\nu_c$). Panel (a): for an ISM medium; Panel (b): for a wind
medium.}\label{Closure_Relation}
\end{figure*}

The standard fireball shocks model (Rees \& {M\'{e}sz\'{a}ros} 1992, 1993;
{M\'{e}sz\'{a}ros} \& Rees 1997; Sari et al. 1998; for reviews, see Zhang \&
{M\'{e}sz\'{a}ros} 2004; Piran 2005; M\'esz\'aros 2006) has been found to
successfully explain the sparse broad-band afterglow data in pre-{\em Swift}
era (e.g. Panaitescu \& Kumar 2002). In the framework of the model, the GRB
central engine powers a relativistic jet that is composed of a series of shells
with variable Lorentz factors. Irregular collisions among these shells produce
the highly variable prompt gamma-rays. As the fireball is decelerated by the
ambient medium, a forward shock propagates into the medium and powers the
long-term broad band afterglows ({M\'{e}sz\'{a}ros} \& Rees 1997; Sari et al.
1998). The decay of the afterglows with time is expected to be a power law with
an index $\sim 1.0$, which would steepen to $1.5 \sim 2$ with an achromatic
break if the fireball is collimated into a conical jet (Rhoads 1999; Sari et
al. 1999). The SPL decay behavior of the 19 SPL XRT lightcurves is consistent
with the prediction of the models. As shown in Paper II, the normal decay phase
in the shallow-to-normal decaying segment is consistent with the closure
relations predicted by the forward shock models (Sari et al. 1998; Chevalier \&
Li 2000; Dai \& Cheng 2001; Zhang \& M\'esz\'aros 2004), favoring the idea that
the X-ray afterglow is of the external shock origin, and that the shallow decay
segment is due to long-lasting energy injection. Figure \ref{Closure_Relation}
presents the model predictions of the closure relations as compared with the
two samples (for the canonical sample data are for the normal decay segment).
It is found that most of the GRBs are roughly consistent with the closure
relations prior to the jet-break for the constant density (ISM) model,
suggesting that these XRT lightcurves might be produced by forward shocks. GRBs
061007 and 080319B are marginally accommodated with the closure relations of
the pre-jet-break wind model and the post-jet-break ISM model. This suggests
that some GRBs may be in a wind-medium (see also Racusin et al. 2008), or with
very narrow jet opening angles (see also Schady et al. 2007). We note again
that not all X-ray afterglows can be interpreted within the forward shock model
due to the chromatic features observed in X-ray/optical lightcurves of some
bursts (Panaitescu et al. 2006; Liang et al. 2007, 2008).

Within the forward shock model, the similar late time luminosity for both SPL and canonical X-ray afterglows
suggest that the total afterglow energetics of the two groups of bursts may be comparable. The difference
then lies in that the SPL GRBs eject the majority of energy promptly with a narrow distribution of large
Lorentz factors, while the canonical GRBs either eject
the same amount of energy over a long period of time or over a wide range of Lorentz factor distribution. The
prompt gamma-ray efficiencies of the two groups of GRB, on the other hand, have to be different: the
canonical GRBs typically have a higher gamma-ray emission efficiency than the SPL ones (Zhang et al. 2007c).

\subsection{The X-rays as an independent emission component prior to the prompt gamma-rays}
\begin{figure*}
\includegraphics[angle=0,scale=0.8]{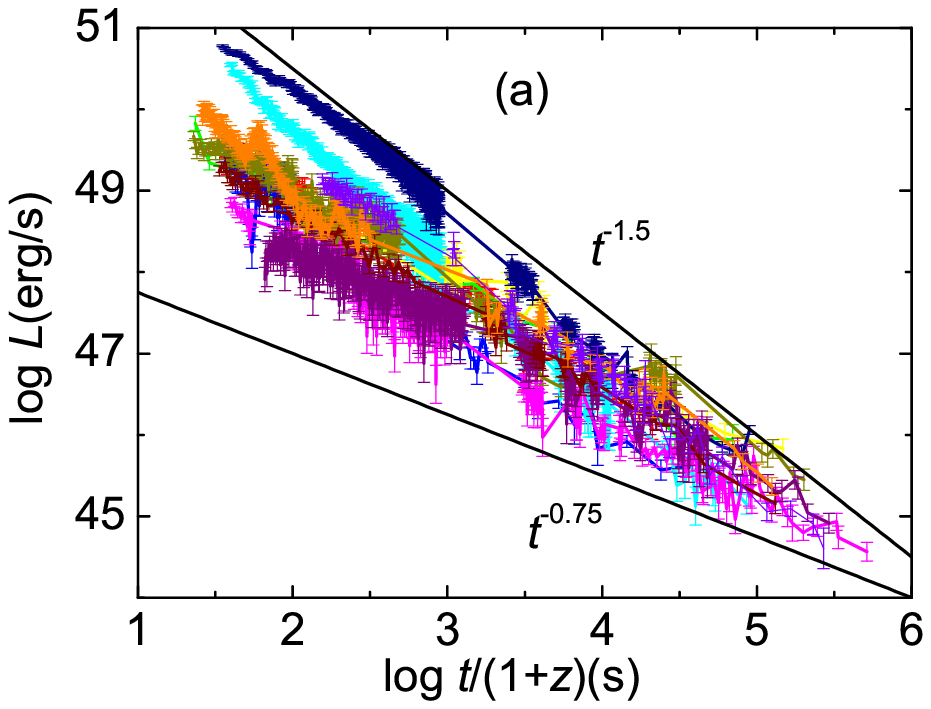}
\includegraphics[angle=0,scale=0.8]{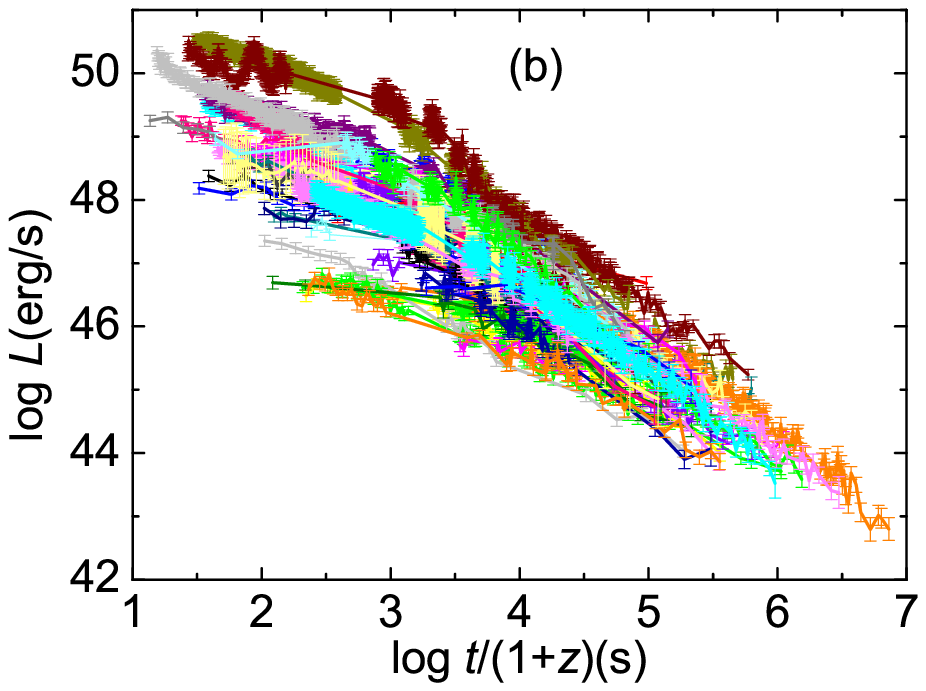}
\includegraphics[angle=0,scale=0.8]{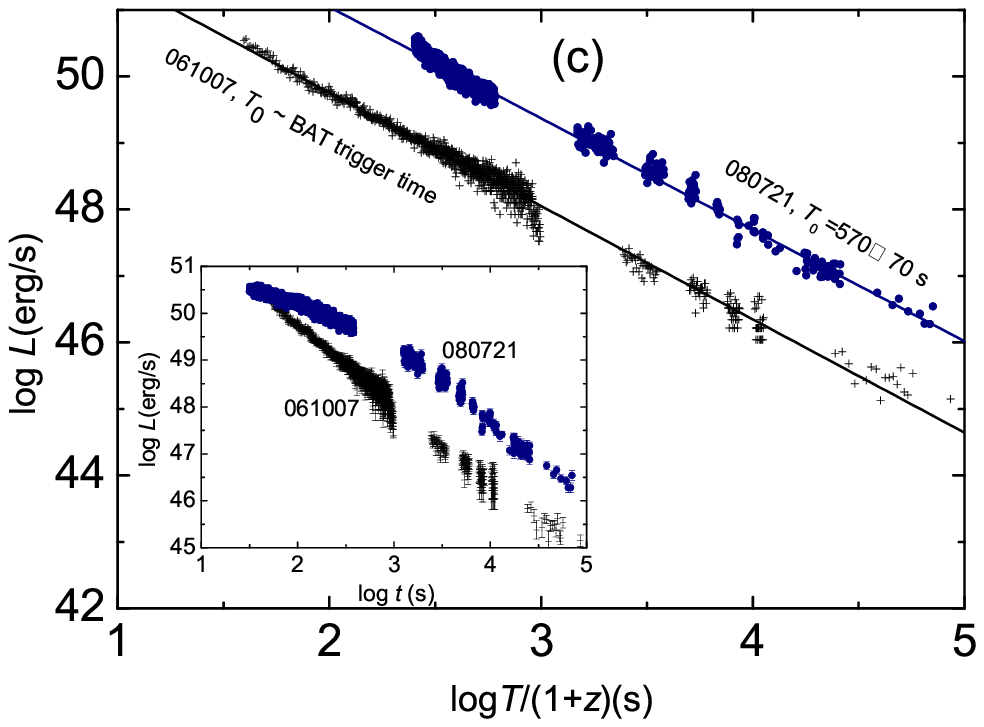}
\hfill
\includegraphics[angle=0,scale=0.8]{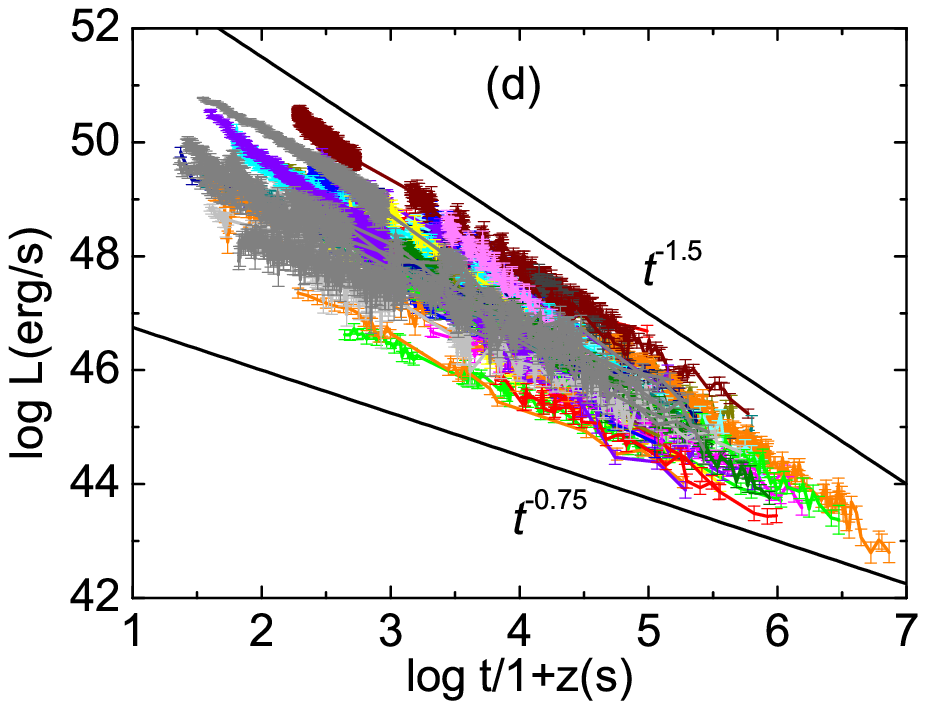}
\caption{X-ray luminosity as a function of time: (a) the observed SPL decay XRT
lightcurves referenced to the BAT trigger times; (b) the observed lightcurves
of the canonical sample referenced to the BAT trigger time; (c) Illustrations
of the XRT lightcurve of GRB 080721 referenced to the $T_0$ and to the BAT
trigger time (inset) with comparisons to the XRT lightcurve of GRB 061007; (d)
comparison of the observed SPL lightcurves (grey) referenced to the BAT trigger
times and canonical XRT lightcurves referenced to shifted $T_0$. The decay
slopes of the lightcurves predicted by external shock models is
$0.75<\alpha_X<1.5$, which are shown as solid lines. } \label{T0_LC}
\end{figure*}

\begin{figure}
\includegraphics[angle=0,scale=0.8]{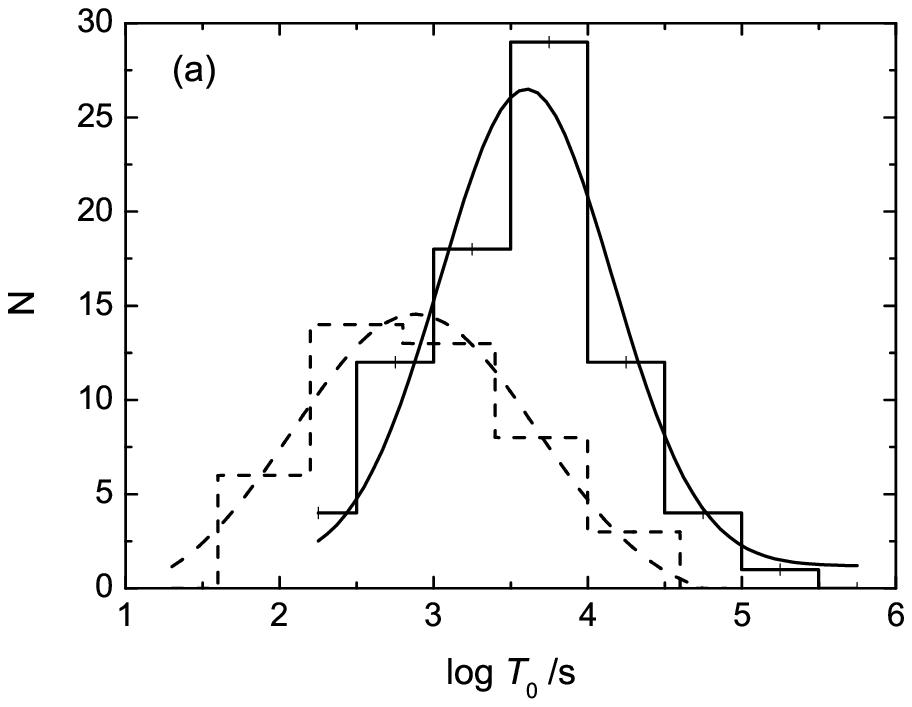}
\includegraphics[angle=0,scale=0.8]{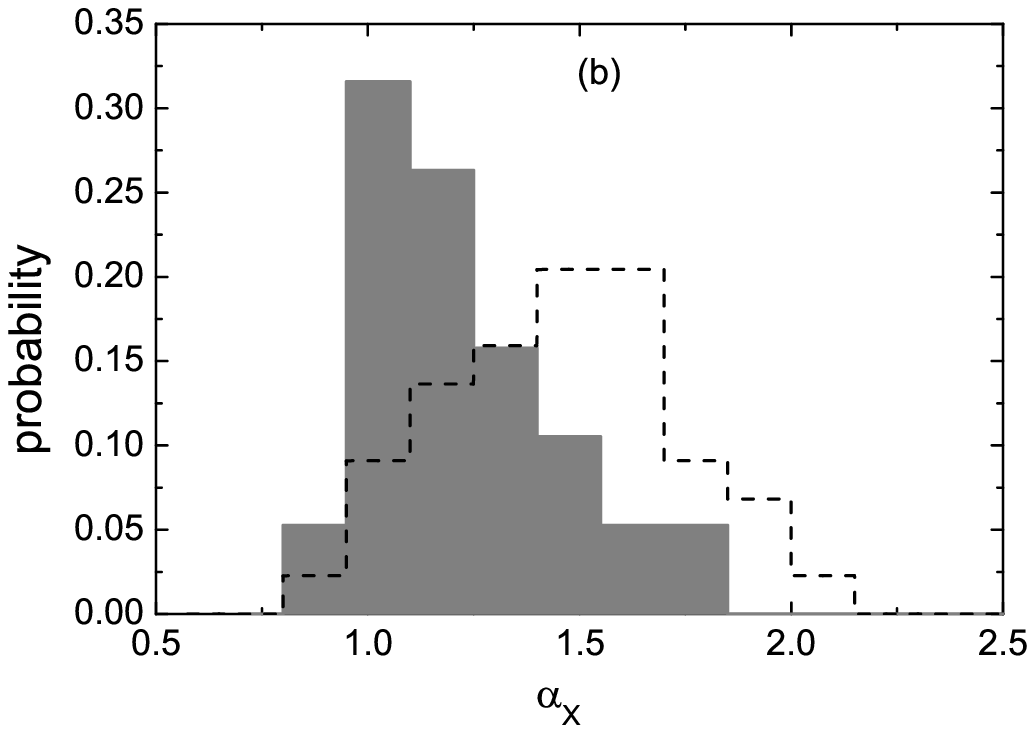}
\caption{{\em Panel (a)}---Distributions of $T_0$ in the observed frame for 80
GRBs ({\em solid line}) and in the burst rest frame for 44 GRBs with redshift
measurements ({\em dashed line}) for the canonical sample. {\em Panel
(b)}---Distribution of the decay slopes of the X-rays with time referenced to
$T_0$ for the 44 GRBs in the canonical sample ({\em dashed line}) in comparison
with the SPL sample ({\em solid line}).} \label{T0_Dis}
\end{figure}

Yamazaki (2009) explained the shallow-to-normal decay behavior of the canonical
X-ray lightcurves as due to the zero time effect. In his model, the X-ray
emission intrinsically decays with a single power law, with starting time point
($-T_0$, with respect to the GRB trigger time) prior to the trigger of the
prompt gamma-rays\footnote{Note that $T_0$ is the onset of the X-ray emission
component, but not the peak time of the X-ray emission (Huang et al. 2002;
Liang et al. 2006; Yamazaki 2009). }. The observed shallow-to-normal decaying
feature is caused by improper zero time point effect. The $T_0$ of the observed
SPL XRT lightcurves would be close to the GRB trigger time. We test this
intriguing possibility by searching for a proper $T_0$ earlier than the BAT
trigger time to make the observed shallow-to-normal decay segment be a single
power law segment. The shallow-to-normal segment in the XRT lightcurves is well
fit with a broken power-law,
\begin{equation}
F=F_0 \left[\left(\frac{t}{t_{\rm b}}\right)^{\omega {\alpha_1}}+\left(\frac{t}{t_{\rm b}}\right)^{\omega
{\alpha_2}}\right]^{-1/\omega},
\end{equation}
where $\omega$ describes the sharpness of the break, which is taken as 3 in this analysis (Liang et al. 2007). Since
$\alpha_2$ is less affected by the $T_0$ effect, we assume that the power-law index of an intrinsic XRT lightcurve is
$\alpha_2$, and search its $T_0$ by fitting the observed shallow-to-normal segment with
\begin{equation}
F=F_0^{'}\left(\frac{t+T_0}{T_0}\right)^{-\alpha_2}.
\end{equation}
As an example, we show the XRT lightcurves of GRB 080721 referenced to BAT
trigger time and to $T_0$ in Fig. \ref{T0_LC}(c), along with the XRT lightcurve
of GRB 061007, the most prominent case in the sample of the SPL XRT
lightcurves. The distributions of $\log T_0$ and the decay slopes of the X-rays
referenced to $T_0$ are shown in Fig. \ref{T0_Dis}. Both the distributions  of
$\log T_0$ in the observed frame and in the burst frame are well fitted with a
Gaussian function, centering at $3.60\pm 0.55\ (1 \sigma)$ and $2.88\pm 0.79\
(1 \sigma)$.  The SPL sample would be those GRBs whose $T_0$ are close to the
BAT trigger time. One might suspect whether the {\em true} $T_0$ distribution
is bimodal by combining the two groups of GRBs. In our sample, the derived
$T_0$ for the canonical sample is longer than 100 seconds. We thus set an upper
limit of 100 seconds for the $T_0$ of X-rays for the SPL sample. Among $\sim
400$ {\em Swift} GRBs with detections of X-ray afterglows, about half have a
shallow-to-normal decay pattern in their XRT lightcurves (see also Evans et al.
2009). Eighty ones are
well-sampled and are selected for our analysis. Considering this sample as a
representative one of the canonical sample, a probability of $5\%$ for
$T_0<100$ is inferred from the derived $T_0$ distribution of the 80 GRBs,
roughly consistent with the percentage of the SPL GRBs in the current Swift GRB
sample, i.e., $19/400$. Therefore, the current data is consistent with the
hypothesis that SPL GRBs are from the same parent sample as the canonical ones,
but belong to the lower end of the $\log T_0$ distribution derived from the
canonical sample.

\begin{figure*}
\includegraphics[angle=0,scale=1]{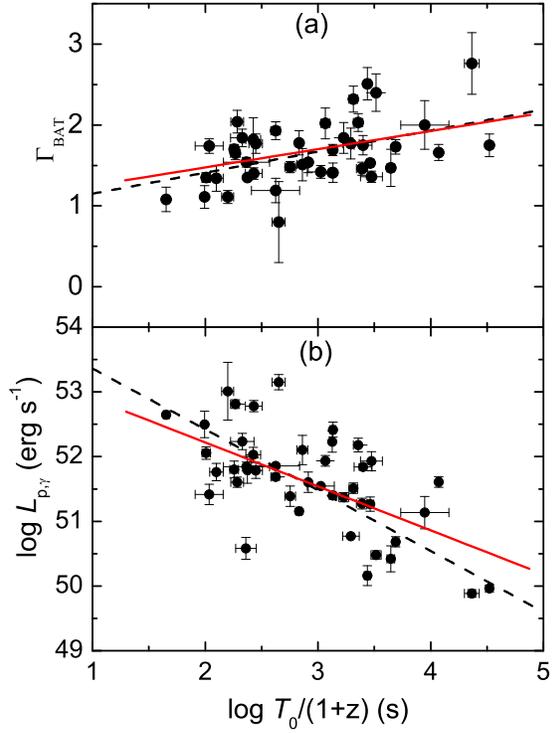} 
\caption{Relations of $T_0$ in the burst frame to the photon index ({\em panel
a}) of prompt gamma-rays and the peak luminosity ({\em panel b}) for the
canonical GRBs in our sample. The dashed and solid lines are the best fit and a
robust fit to the data.} \label{T0_Liso}
\end{figure*}

The $T_0$ is the time interval between the starting time of the X-ray emission
and that of the prompt gamma-ray emission. It is worth investigating if there
are any correlations between $T_0$ and some observables of the prompt
gamma-rays. We find that $T_0$ is not correlated with $T_{90}$, but it is
tentatively anti-correlated with the isotropic gamma-ray peak luminosity (or
isotropic gamma-ray energy) and the photon index of the prompt gamma-rays. As
shown in Fig. \ref{T0_Liso}, a $L_p-T_0/(1+z)$ correlation and a $\Gamma_{\rm
BAT}-T_0/(1+z)$ correlation with the Spearman correlation coefficients
$r=-0.58$ (chance probability $p<10^{-4}$) and $r=-0.39$ ( $p\sim 10^{-2}$) are
derived. This implies that brighter or harder bursts tend to have a shorter
time interval between the burst itself and the prior X-ray emission. However,
the correlations are not tight, and should be taken with caution.

Precursors at tens or even hundreds of seconds prior to main bursts have been
detected in some GRBs (e.g. Lazzati 2005; Burlon et al. 2008). It would be
interesting to test whether these precursors are related to the prior emission
discussed in this paper. Well-sampled precursors were detected in GRBs 060124
and 061121 in the Swift GRB sample. Their XRT lightcurves also behave as
canonical ones. We therefore check their inferred $T_0$ and consistency with
the leading time between the precursor and the main burst. We find that $T_0$
leads the precursor time significantly. GRB 060124 has a precursor $\sim 570$
seconds prior to the main burst peak (Romano et al. 2006), but its X-ray $T_0$
is $(5.57\pm 0.58)\times 10^3$ seconds before the trigger (precursor), which is
$(6.14\pm 0.58)\times 10^3$ seconds before the main burst. The precursor of GRB
061121 is at $\sim 60$ seconds prior to the main burst(Page et al. 2007), but
its X-rays $T_0$ is $(3.09\pm 0.08)\times 10^3$ seconds prior to the GRB
trigger (precursor) and $(3.15 \pm 0.08)\times 10^3$ seconds before the main
burst. This seems to suggest that the prior emission is another component other
than the precursor. However, we emphasize that $T_0$ is the beginning of the
prior emission component, not necessarily the peak of the prior emission, which
is expected to be later from $T_0$ and closer to the GRB main peak. The
possibility that the precursor is related to the prior emission is not ruled
out. Since the $T_0$'s derived in this paper are all with respect to the BAT
trigger time, it may pick up some precursors rather than the main bursts as the
reference point. This would bring confusions to the measured $T_0$ and
contribute the scatter of the correlations presented in Fig. \ref{T0_Liso}.

In general, the $T_0$ effect is a major issue of presenting lightcurves in the
log-log space. Shifting $T_0$ can effectively modify the power-law decay
indices of a lightcurve, which directly affect the theoretical interpretation
of the phenomena. Recalling the history of the GRB study, we caution that $T_0$
is a two-edged weapon that can be used equally for good or bad. Moving $T_0$
after the GRB trigger time, Liang et al. (2006) found that the early steep
decay segment of the XRT lightcurves can be explained by the tail emission of
the last emission epoch, and that X-ray flares are consistent with being late
internal emission related to the central engine. On the other hand, by setting
$T_0$ to near the X-ray flares, Piro et al. (2005) found that the X-ray flares
in GRBs 011121 and 011211 are consistent with the onsets of the X-ray
afterglows, and hence, missed to report the first detections of flares in
GRBs\footnote{Note that X-ray flares are likely related to reactivation of the
central engine at later times. time should be set to before the onset of the
X-ray flares (Liang et al. 2006). They are an independent component
superimposed onto the underlying prior X-ray emission component.
Phenomenologically, if one plots X-ray afterglow lightcurve with respect to
$T_0$, then early X-ray flares (occurring at epochs shorter than $T_0$ after
the trigger time) would appear ``narrower", i.e., having steeper rising and
falling indices. The late X-ray flares (those occurring at epochs longer than
$T_0$ after the trigger time) would look similar to the ones plotted
referencing the trigger time.}. Here, by putting $T_0$ prior to the GRB trigger
time, we argue that a shallow-to-normal decay segment can become a single
power-law (Yamazaki 2009). Such a suggestion may be at risk. However, since it
is an intriguing possibility and can be tested by future observations, in the
following we will explore the consequence of such an assumption. We will
investigate whether moving $T_0$ would make a better consistency between the
canonical and the SPL samples.

\begin{figure}
\includegraphics[angle=0,scale=0.8]{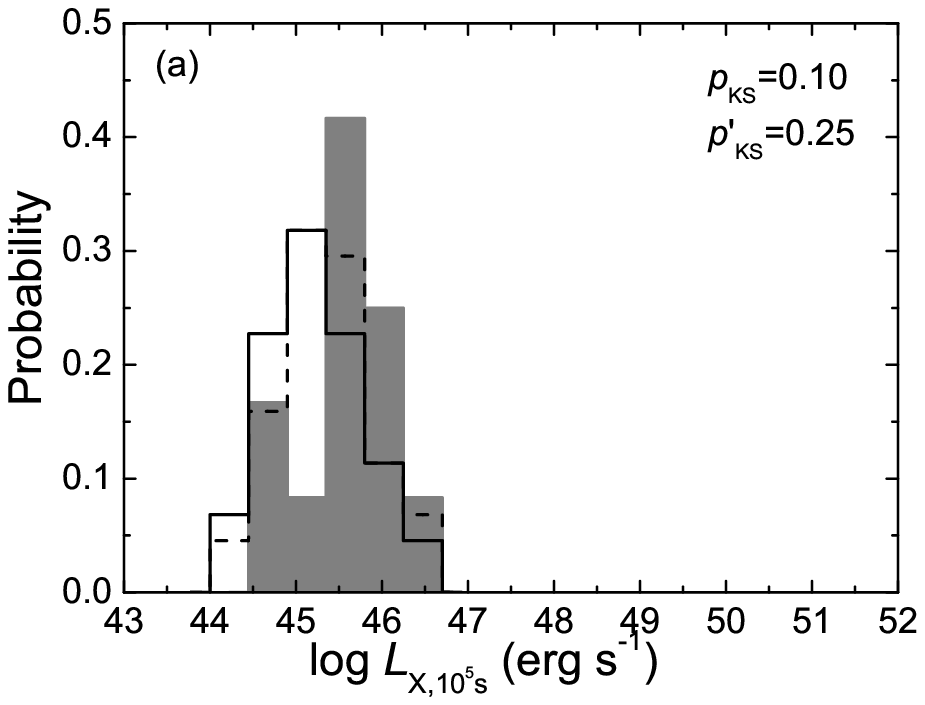} \\ 
\hfill
\includegraphics[angle=0,scale=0.8]{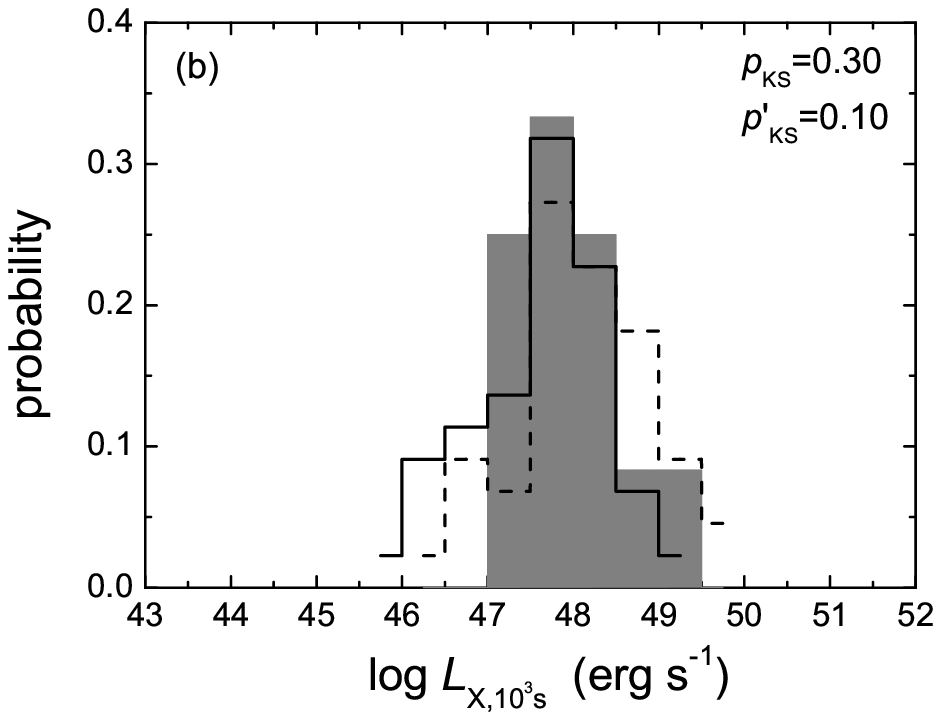} \\ 
\hfill
\includegraphics[angle=0,scale=0.8]{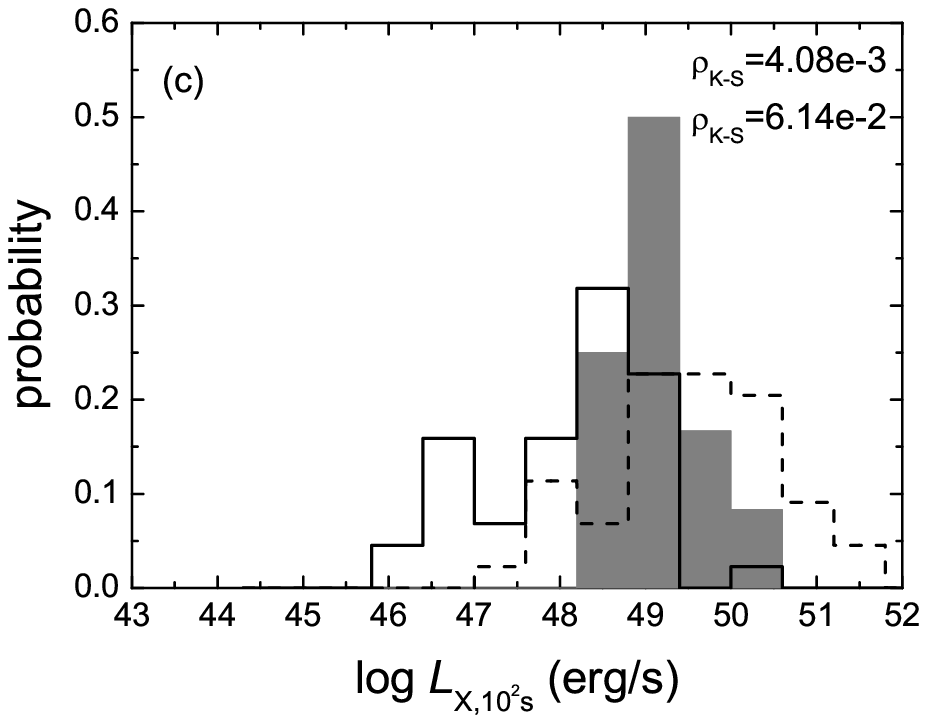} 

\caption{Luminosity distributions at $10^5$, $10^3$, and $10^2$ seconds with
respect to the BAT trigger times ({\em solid lines}) and with respect to the
$T_0$'s ({\em solid lines}) for the canonical sample, as compared with the SPL
sample ({\em grey columns}).} \label{Luminsoity_Dis}
\end{figure}

All XRT lightcurves referenced to $T_0$ are shown in Fig. \ref{T0_LC}(d) along
with the observed SPL XRT lightcurves. It is interesting that they well trace
the observed SPL XRT lightcurves. The distributions of the X-ray luminosity at
$t=10^2$, $10^{3}$, and $10^{5}$ seconds referenced to $T_0$ ($L^{T_0}_{X, t}$)
and to the BAT trigger time ($L^{\rm BAT}_{X, t}$) are shown in Fig.
\ref{Luminsoity_Dis}, with comparisons to the SPL GRBs\footnote{With a sample
of 16 pre-{\em Swift} X-ray afterglow lightcurves, Gendre \& Bo\"{e}r (2005)
suggested two classes of GRBs defined by the X-ray afterglow luminosity. This
signature is possibly related to the two groups of XRT lightcurves discussed in
this paper. Although the early X-ray luminosities (at $10^2$ second) of the SPL
XRT lightcurves tend to be brighter than those of the canonical ones, we do not
confirm the bimodal distributions of the X-ray luminosity at $10^3$ second and
$10^5$ second, as shown in Fig.\ref{Luminsoity_Dis}.}. It is found that
$\log L^{\rm BAT}_{X, {10^2}s}$ falls in the range of [$46$, $50$] with an
average of $48.1\pm 0.9$(1$\sigma$), and $\log L^{\rm T_0}_{X, {10^2}s}$ in the
range of [$47$, $52$] with an average of $49.5\pm 1.0$(1$\sigma$), typically
being larger than $\log L^{\rm BAT}_{X, {10^2}s}$ with 1.4 orders of magnitude.
For the SPL GRB sample, the $\log L^{\rm SPL}_{X, {10^2}}$ distribution is in
the range of [$48$, $51$] with an average of $49.8\pm 0.5$(1$\sigma$), roughly
consistent with $\log L^{\rm T_0}_{X, {10^2}s}$. We measure the consistency of
the luminosity distributions to that of the SPL GRBs by the K-S test, which
gives $p_{K-S}=4.08 \times 10^{-3}$ and $p_{K-S}=6.14 \times 10^{-2}$ for $\log
L^{\rm BAT}_{X, {10^2}s}$ and  $\log L^{\rm T_0}_{X, {10^2}s}$, respectively,
indicating that the distribution of $\log L^{\rm T_0}_{X, {10^2}s}$ is more
consistent with that of the SPL GRB sample. The distributions of $L^{\rm BAT}_{X,
{10^5}s}$ and $L^{\rm T_0}_{X, {10^5}s}$ are consistent with each other, and
they also well agree with that of the SPL GRB sample.

\begin{figure}
\includegraphics[angle=0,scale=0.7]{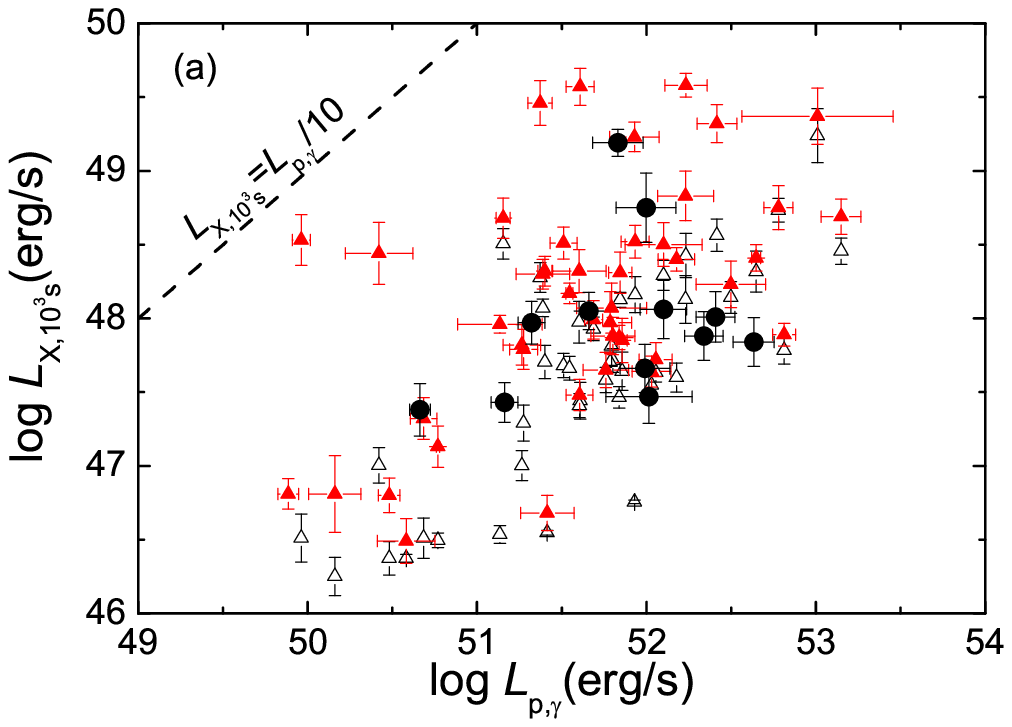}
\includegraphics[angle=0,scale=0.7]{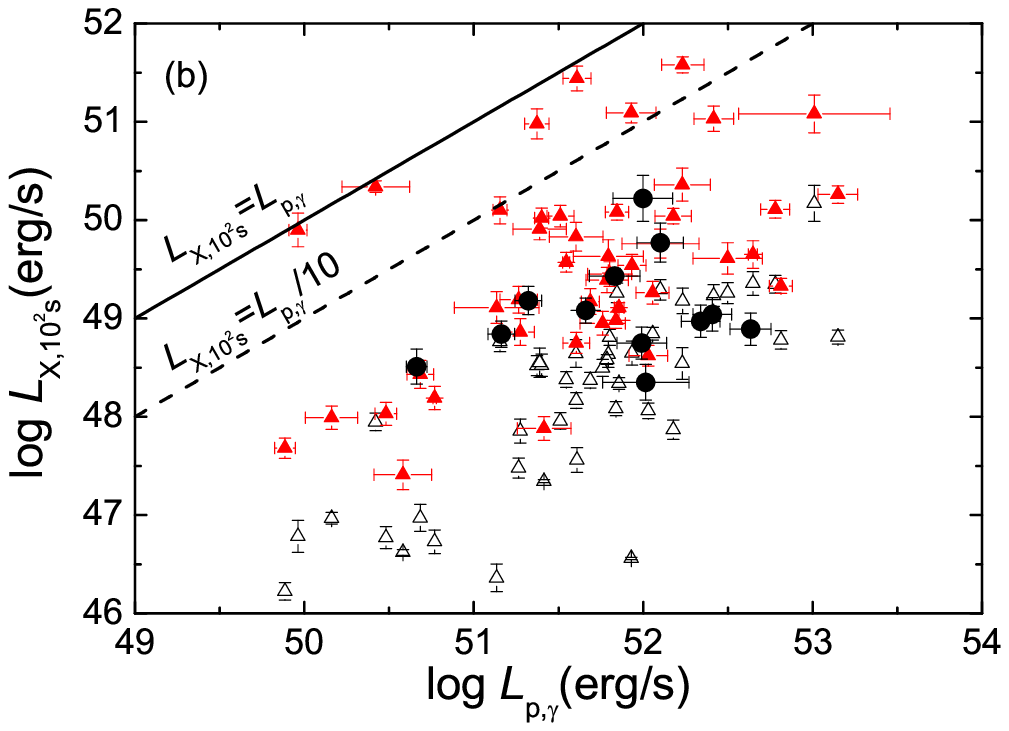}
\includegraphics[angle=0,scale=0.7]{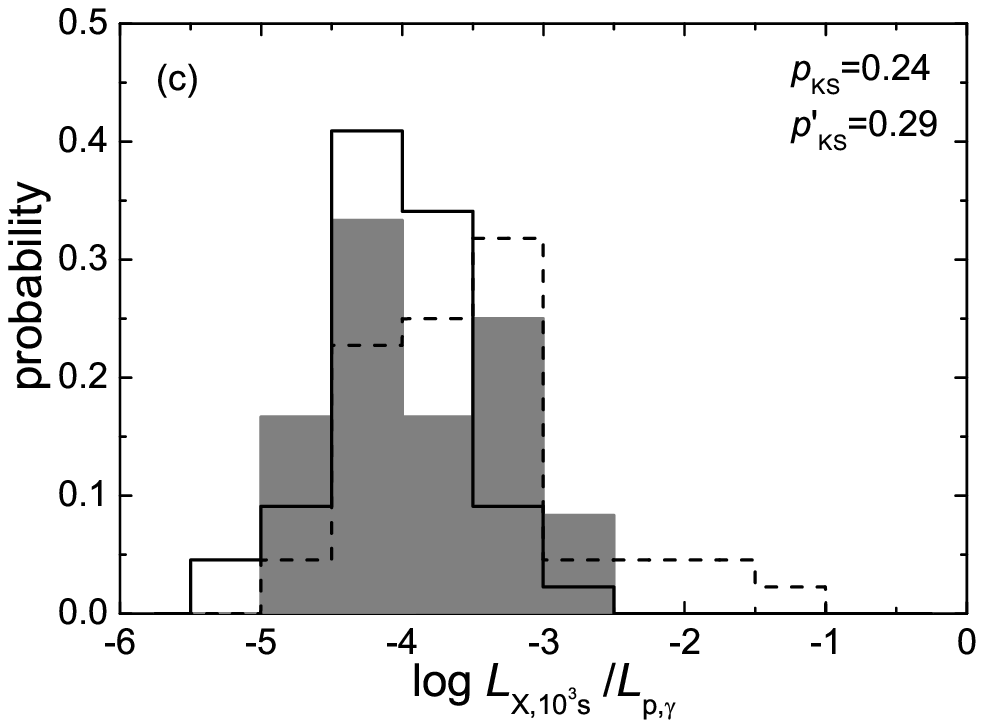}\hfill
\includegraphics[angle=0,scale=0.7]{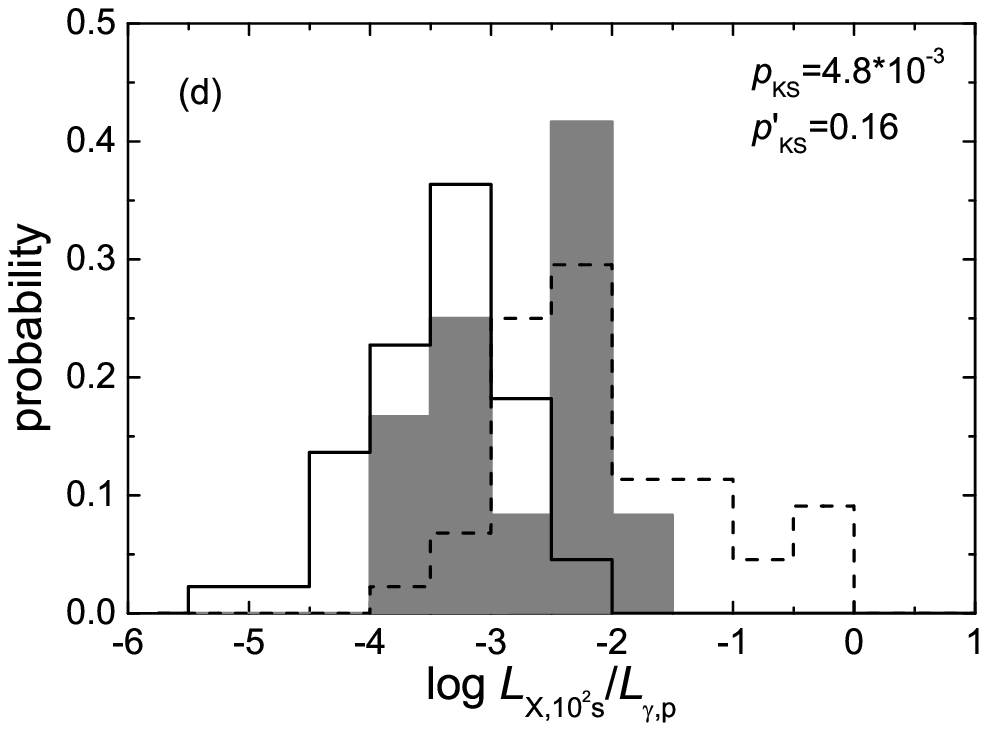}

\caption{{\em Panels (a) and (b)} --- Comparisons of the SPL sample ({\em solid
circles}) with the canonical GRB samples in the ($L_{\rm X, t}-L_{\gamma,
p}$)-planes, with $t=10^2$ seconds and $10^3$ seconds referenced to the BAT
trigger time ({\em open triangles}) or to $T_0$ ({\em solid triangles}). The
lines are $L_{\rm X,t}=L_{\gamma,p}$ and $L_{\rm X,t}=0.1L_{\gamma,p}$. {\em
Panels (c) and (d)} --- Distributions of $\log L^{\rm BAT}_{\rm X}/L_{\gamma,
p}$ ({\em solid line}) and $\log L^{T_0}_{X}/L_{\gamma, p}$ ({\em dashed line})
at $10^2$ ({\em panel c}) and $10^3$ ({\em panel d}) seconds for the canonical
sample with comparisons to the SPL sample ({\em grey columns}).}\label{LX-Lg}
\end{figure}

In order to compare the X-ray luminosity with the peak luminosity of prompt
gamma-ray emission $L_{p,\gamma}$, we show the SPL GRBs ({\em solid circles})
and canonical GRB samples in the ($L_{X, 10^2s}-L_{\gamma, p}$)-plane and
($L_{X, 10^3s}-L_{\gamma, p}$)-plane, where the time of the X-ray luminosity
for the canonical samples is referenced to the BAT trigger time or to $T_0$ in
Fig. \ref{LX-Lg}. Comparisons of the distributions of $\log
L^{BAT}_{X}/L_{\gamma, p}$ and $\log L^{T_0}_{X}/L_{\gamma, p}$ at $10^2$ and
$10^3$  seconds for the canonical sample to the SPL sample are also shown in
Fig. \ref{LX-Lg}. Except for seven GRBs, the $L^{\rm T_0}_{X, {10^2}s}$ of the
other GRBs are smaller than $0.1L_{p,\gamma}$, and the $L^{\rm T_0}_{X,
{10^3}s}$ of all GRBs are smaller than $0.1L_{p,\gamma}$. For the seven GRBs,
their $ L^{\rm T_0}_{X, {10^2}s}$ are comparable to $L_{p,\gamma}$. It is
possible that the peak time of the X-ray emission is at a time later than 100
seconds with respect to $T_0$, and their $ L^{\rm T_0}_{X, {10^2}s}$ would be
over-estimated. Excluding the seven GRBs, more consistency is observed between
the two groups of GRBs in the $L_{p,\gamma}-L^{T_0}_{X}$ planes. Even
considering the seven cases, the $L^{T_0}_{X,10^2}$ of the canonical sample is
still more consistent with the SPL sample than $L^{BAT}_{X,10^2}$(testing by
K-S test, as marked in Fig. \ref{LX-Lg}).

The analysis above indicates that the canonical XRT lightcurves may have the same physical origin as the SPL XRT
lightcurves, and the apparent shallow-to-normal segment is due to improper zero time point. The zero time points of
those observed single power-law XRT lightcurves are possibly comparable to the BAT trigger time. The fraction of these
GRBs is very small in the current {\em Swift} GRB sample, i.e., 19 cases out of $\sim 400$ GRBs. The X-rays may be a
long-lasting emission component prior to the GRB trigger time as suggested by Yamazaki (2009). The discovery of the
X-ray flares following a good fraction of GRBs suggest that a long-live GRB central engine is common for GRBs (Burrows
et al. 2005; Zhang et al. 2006; Fan \& Wei 2005; King et al. 2005; Dai et al. 2006; Perna et al. 2006; Proga \& Zhang
2006). The prior emission requires that the central engine activity time scale is stretched even longer.

In principle, a prior X-ray emission decaying as a single power-law may be originated both from an external shock or
central-engine-powered internal dissipation. In the scenario of a central-engine-powered X-ray emission model, the
X-rays might be powered by an unknown internal dissipation mechanism, and the X-ray luminosity is conjectured to track
the accretion power at the central engine. The accretion rate by the central engine may be expressed by (e.g. Kumar et
al. 2008)
\begin{equation}\label{Mdot}
\dot{M}\sim \dot{M}(t_0) (1+\frac{3}{2s+1}\frac{t-t_0}{t_{acc}})^{-4(s+1)/3},
\end{equation}
where $t_{acc}$ is a characteristic timescale of accretion and $0<s<1$. The
luminosity can be then estimated by $L=\eta\dot{M}c^2$. The observed decay
slope of the X-rays is $1\sim 2$, being roughly consistent with Eq. \ref{Mdot}.
The parameter $s$ is quite uncertain. As shown in Fig. \ref{T0_LC}, the decay
slopes of the lightcurves are $-4/3\sim -5/3$. Critical concern on this
scenario is that it requires the central engine to be active as long as $10^5$
seconds, even $10^7$ seconds (e.g., GRB 060729). The strong dependence of the
neutrino annihilation mechanism on the mass accretion rate makes it difficult
to explain the data (e.g., Barkov \& Komissarov 2009). The neutrino mechanism
requires the mass accretion rate to stay over few $0.01M_\odot$/s (e.g., Fan,
Zhang \& Proga 2005; Popham et al. 1999). Long-lasting accretion implies the
progenitor mass is above hundreds of solar mass. However, the mass of a WR star
is 9-25 $M_\odot$, though some observations suggested that it can be as high as
83 $M_\odot$(Schweickhardt et al. 1999; Crowther 2007).

A more straightforward model to explain the power-law decay of the X-rays is the
external shock model. As shown in Fig. \ref{Closure_Relation}, the X-rays are
generally consistent with the external shock models, similar to the GRB
afterglows. They may be also produced by interaction of an early ejecta
launched prior to the formation of the GRB jet with the circumburst medium. The observed
flux should increase and then decay as a single power law post the peak of the
emission. Within the ISM forward shock model the decay slopes of the X-rays
varies between 0.75 and 1.5, depending on the observed spectral index and
medium properties (e.g., paper III). We thus mark $L\propto t^{-0.75}$ and
$L\propto t^{-1.5}$ in Fig. \ref{T0_LC}. It is found that the data are well
consistent with the model prediction. Note that the $T_0$ is the start time of
the X-ray emission component, but not its peak time.  In the external shock
scenario, the peak time of the X-rays can be estimated by (M\'esz\'aros \& Rees
1993; Sari \& Piran 1999)
\begin{equation}
t_p\approx 10^3(1+z)E_{iso,X,52}^{1/3}\eta_{0.2}^{-1/3}n^{-1/3}\Gamma_{0,2}^{-8/3},
\end{equation}
where $E_{iso, X,52}=E_{iso, X}/10^{52}{\rm erg}$, $\eta_{0.2}=\eta/0.2$ is the
radiative efficiency, $n$ is the medium density, and
$\Gamma_{0,2}=\Gamma_0/10^2$ is the initial Lorentz factor of the fireball. As
shown in Fig. \ref{T0_Dis}(a), the $T_0$ of the X-rays for most bursts are
several thousands of seconds. This suggests that the peak time of the X-rays
for some GRBs may be close to the BAT trigger time.

One related question is the jet break in the prior emission. Inspecting the SPL
sample (Fig. \ref{XRT_LC}), we find that only GRBs 080413 and 080913B show a
jet-like break at $\times 10^5$ seconds, and that the others have no break
feature. For the canonical sample, our $T_0$ search can make the shallow-decay
phase to have the same decay slope as the normal decay phase. However, if there
is a late jet-like break in the canonical lightcurve, such a break should still
exist after the $T_0$ shift. An example is GRB 060729 (Fig. \ref{T0_LC}c). In
general these breaks tend to be late, too. These facts imply that the prior
X-ray emission would be usually from a jet with a large opening angle. Again
taking GRB 060729 as an example. The X-ray emission was observed up to 642 days
after the GRB trigger and a jet-like break was observed at around 1 year after
the GRB trigger (Grupe et al. 2009)\footnote{The steep temporal decay and
significant spectral softening after the break also favor a spectral origin for
the break(Grupe et al. 2009).}. A jet-like break is usually seen in the
lightcurves of the optical afterglows at days to one week post the GRB
triggers. The jet that is associated with the prompt gamma-ray emission thus
might be generally narrower than the jet for the prior X-ray emission, if the
optical emission is the afterglows of the jet related to the prompt emission.

\section{CONCLUSIONS}
By systematically analyzing the XRT lightcurves for $\sim 400$ {\em Swift} GRBs
detected by June 2009, we have investigated the properties of the GRBs with a
SPL decaying XRT lightcurve and made comparisons between these GRBs (the SPL
sample) and the GRBs having a canonical XRT lightcurve (the canonical sample).
We only find 19 GRBs whose XRT lightcurves decay with a SPL from tens to $\sim
10^{5}$ seconds post the GRB triggers. The decay slopes of these lightcurves
range from $1-1.7$. The fraction of these GRBs in the whole {\em Swift} sample
is small, i.e., 19 out of  $\sim 400$, suggesting that the SPL lightcurves are
much less common than the canonical ones. There is no statistical difference
between the distributions of $T_{90}$, $E_{iso,\ \gamma}$, and $\Gamma_{\rm
BAT}$ of the prompt gamma-ray parameters for the two groups (SPL vs. canonical)
of GRBs. No significant spectral evolution is observed for the X-rays with the
SPL decay, similar to that observed in the canonical GRBs (Paper II). No
statistical difference of the X-ray spectra is found between the two groups of
GRBs, and the power-law indices of the $E_{\rm iso, X}-E_{\rm iso, \gamma}$
relations for the two GRB samples are almost the same. However, the SPL sample
tends to have a slightly lower neutral hydrogen absorption column by the host
galaxy and a slightly larger energy release in the X-ray band than the
canonical sample.

The SPL lightcurves in burst frame gradually merge into a conflux at around one day post the GRB trigger. A Gaussian
function fit in logarithmic scale to the luminosity distribution at $10^5$ s yields $\log (L_{10^5{\rm cm}}/{\rm
ergs~s^{-1}})=45.50\pm 0.70$. The normal decay phase in the shallow-to-normal segment of the canonical GRBs share the
similar feature. Confronting the data with the predictions of the external shock models, we find that the SPL
lightcurves are generally consistent with the models, similar to the normal decay phase of the canonical lightcurves.
The external shock origin of both the SPL lightcurves and the normal decay segments in the canonical XRT lightcurves
is favored. If the shallow decay is due to energy injection into the fireball, the total energy budget
after injection for both samples of GRBs is similar.

The apparent shallow decay phase in the canonical sample may also be due to the
$T_0$ effect of a SPL X-ray emission component prior to the GRB trigger, as
suggested by Yamazaki (2009). By setting the $T_0$ at an epoch before the GRB
trigger, we show that a shallow-to-normal segment becomes a SPL with the same
decay index as the normal decay phase. The XRT lightcurves referenced to $T_0$
also well trace the observed SPL XRT lightcurves. This likely suggests that the
X-rays might be a long-lasting emission component starting before the GRB
trigger and the SPL sample might be composed of the GRBs whose $T_0$'s are
close to the GRB trigger times. Although precursors at tens or even hundreds of
seconds prior to main bursts were detected in some GRBs, their spectral
properties reveals that the precursors are not a phenomenon distinct from the
main event, and the prior time offset of the X-rays in our analysis is much
larger than that the offset of the known precursors. However, the peak time of
the prior emission is different from $T_0$, so the possibility that the
precursors are related to the prior emission is not ruled out.

As shown above the external shock model invoking an emission component prior to
the GRB prompt emission may give a unified picture to interpret the canonical
and SPL XRT lightcurves. In this scenario the GRB phenomenon should invoke two
different ejecta components that are responsible for the observed X-ray
afterglow and the prompt emission (along with the optical afterglow in most
cases), respectively. The mixture of radiation from these two components would
bring complication and confusion on the identifications of them. The
well-sampled lightcurves of early optical/IR afterglows usually show a smooth
onset feature as expected from the deceleration of the GRB fireball for some
GRBs, such as GRBs 060418 and 060607A (Molinari et al. 2007), and a jet break
is also usually observed in the late optical lightcurves (Sari et al. 1999).
These facts suggest that the optical afterglows would be dominated by the
external shock of the jet that is associated with the prompt emission. On the
other hand, such an early onset feature is not common in XRT lightcurves, only
detected in a few GRBs,  such as GRB 080307 (Page et al. 2009). This is
partially due to the contamination of the X-rays associated with the prompt
emission (e.g. the steep decay tail), but is also consistent with the
hypothesis of an earlier onset of X-ray afterglow. The deficiency of X-ray jet
breaks (Liang et al. 2008; Racusin et al. 2009) is consistent with the
hypothetical wider opening angle of the prior jet. The chromatic breaks in the
optical and X-ray lightcurves for some GRBs (Papers II and III) demand that the
X-ray and optical are two different emission components. The prior emission
model requires that the X-rays should be dominated by the prior emission
component, while the optical emission (in most cases) is related to the ejecta
from the prompt emission. This requires that the X-ray afterglow associated
with the prompt emission jet is buried beneath the X-ray emission related to
the prior component, while the optical emission of the prior component is
outshone by that of the prompt emission component. This is great concern of
this model, along with any other models that invoke two emission components
(e.g. Ghisellini et al. 2008). Detailed theoretical modeling is called for and
is in plan.

\acknowledgments We acknowledge the use of the public data from the Swift data
archive. We appreciate valuable comments/suggentions from the referee. We also
thank helpful discussion with Zi-Gao Dai, Kunihito Ioka, Wei-Hua Lei, Tong Liu,
Milhail Medvedev, Paul O'Brien, Rob Preece, Ryo Yamazaki, Ding-Xiong Wang,
Xiang-Yu Wang, Jian-Yan Wei, and Shuang-Nan Zhang. This work was supported by
the National Natural Science Foundation of China under grants No. 10873002, the
National Basic Research Program (''973" Program) of China (Grant 2009CB824800),
the research foundation of Guangxi University(M30520), and NASA NNG05GB67G,
NNX08AN24G, and NAX08AE57A. BBZ acknowledges the President's Fellow Ship and
GPSA awards from UNLV.


\begin{deluxetable}{llllllllllllll}

\rotate \tablewidth{620pt} \tabletypesize{\footnotesize}
\tablecaption{The observations and our fits for the GRBs with a single power
law decaying XRT lightcurve}

\tablenum{1} \tablehead{ \colhead{GRB}& \colhead{$T_{90}$(s)}&
\colhead{$\Gamma_\gamma$\tablenotemark{a}}&
\colhead{$S_\gamma$\tablenotemark{a}}& \colhead{$t_{1}\sim
t_{2}$(ks)\tablenotemark{b}}& \colhead{$\alpha_x$\tablenotemark{b}}&
\colhead{$\chi^2/$dof\tablenotemark{b}}& \colhead{$\beta_x$\tablenotemark{b}}&
\colhead{$S_x$\tablenotemark{c}}& \colhead{$N_{H}$\tablenotemark{b}}&
\colhead{$z$\tablenotemark{d}}} \startdata
050721&98.4&1.85$\pm$0.19&3.62$\pm$0.32&0.20$\sim$257.24&1.24$\pm$0.02&168/131&$1.43^{+0.33}_{-0.44}$&0.91$\pm$0.22&$12.5^{+7.4}_{-12.3}$&$\ldots$\\
050922C&4.5&1.37$\pm$0.06&1.62$\pm$0.05&0.12$\sim$67.53&1.12$\pm$0.01&201/142&$1.14^{+0.06}_{-0.08}$&0.67$\pm$0.11&$25.0^{+14.9}_{-8.5}$&$2.198^{(1)}$\\
060111B&58.8&0.96$\pm$0.17&1.60$\pm$0.14&0.10$\sim$71.70&1.09$\pm$0.03&122/84&$1.29^{+0.16}_{-0.18}$&0.25$\pm$0.07&$22.0^{+4.7}_{-5.7}$&$\ldots$\\
060116&105.9&1.43$\pm$0.19&2.41$\pm$0.26&0.18$\sim$529.59&1.06$\pm$0.03&11/11&$0.89^{+0.17}_{-0.22}$&0.11$\pm$0.04&$57.4^{+7.4}_{-14.7}$&$6.6^{(2)}$\\
060512&8.5&2.49$\pm$0.32&0.23$\pm$0.04&0.11$\sim$104.01&1.39$\pm$0.03&31/31&$0.93^{+0.18}_{-0.18}$&0.53$\pm$0.15&$<2.35$&$0.4428/2.1^{(3,4)}$\\
061007&75.3&1.0$\pm$0.03&44.41$\pm$0.56&0.09$\sim$194.49&1.71$\pm$0.01&1667/1133&$0.99^{+0.08}_{-0.08}$&17.78$\pm$1.31&$54.4^{+9.6}_{-9.1}$&$1.261^{(5)}$\\
070318&74.6&1.43$\pm$0.09&2.48$\pm$0.11&0.07$\sim$943.92&1.11$\pm$0.01&359/301&$0.82^{+0.03}_{-0.06}$&1.11$\pm$0.16&$56.4^{+3.6}_{-5.4}$&$0.836^{(6)}$\\
070330&9&2.26$\pm$0.27&0.18$\pm$0.03&0.08$\sim$167.03&1.01$\pm$0.03&62/45&$0.85^{+0.07}_{-0.10}$&0.16$\pm$0.06&$6.8^{+3.3}_{-2.2}$&$\ldots$\\
070411&121.5&1.76$\pm$0.11&2.71$\pm$0.16&0.47$\sim$582.08&1.00$\pm$0.04&93/54&$1.18^{+0.19}_{-0.09}$&0.47$\pm$0.23&$145.8^{+152.5}_{-124.0}$&$2.954^{(7)}$\\
071020&4.2&1.11$\pm$0.05&2.30$\pm$0.10&0.07$\sim$632.97&1.03$\pm$0.01&458/370&$1.20^{+0.07}_{-0.12}$&0.86$\pm$0.12&$13.7^{+1.4}_{-3.1}$&$2.142^{(8)}$\\
071025&109&1.72$\pm$0.06&6.20$\pm$0.20&0.15$\sim$316.28&1.47$\pm$0.01&467/306&$1.23^{+0.09}_{-0.08}$&4.42$\pm$0.65&$8.2^{+1.9}_{-1.9}$&$\ldots$\\
080319B&50&1.08$\pm$0.02&81.0$\pm$1.01&0.07$\sim$174.27&1.58$\pm$0.05&1404/978&$0.79^{+0.04}_{-0.04}$&98.82$\pm$6.43&$11.1^{+2.7}_{-2.6}$&$0.937^{(9)}$\\
080413B&8&1.26$\pm$0.27&3.20$\pm$0.10&0.14$\sim$612.23&0.91$\pm$0.01&868/549&$0.90^{+0.05}_{-0.05}$&0.74$\pm$0.10&$24.1^{+4.3}_{-4.2}$&$1.1^{(10)}$\\
080714&33&1.52$\pm$0.08&2.50$\pm$0.10&0.09$\sim$365.64&1.11$\pm$0.02&14/13&$0.61^{+0.18}_{-0.20}$&0.36$\pm$0.07&$9.4^{+14.5}_{-9.4}$&$\ldots$\\
080804&34&1.19$\pm$0.09&3.60$\pm$0.20&0.11$\sim$421.02&1.10$\pm$0.01&75/101&$0.86^{+0.09}_{-0.06}$&0.42$\pm$0.05&$28.2^{+18.9}_{-9.8}$&$2.2045^{(11)}$\\
080906&147&1.59$\pm$0.09&3.50$\pm$0.20&0.08$\sim$304.75&1.19$\pm$0.01&370/250&$1.05^{+0.07}_{-0.16}$&1.23$\pm$0.19&$11.9^{+5.0}_{-4.8}$&$2^{(12)}$\\
090102&27&1.36$\pm$0.08&6.80$\pm$0.03&0.39$\sim$688.48&1.35$\pm$0.01&196/144&$0.86^{+0.08}_{-0.08}$&3.95$\pm$0.61&$64.5^{+14.2}_{-8.6}$&$1.547^{(13)}$\\
090123&131&1.74$\pm$0.12&2.90$\pm$0.02&0.11$\sim$183.64&1.51$\pm$0.02&213/141&$0.79^{+0.14}_{-0.08}$&1.09$\pm$0.24&$0.05^{+2.6}_{-0.1}$&$\ldots$\\
090401B&183&1.37$\pm$0.05&$\sim $10.0&0.08$\sim$782.30&1.36$\pm$0.01&1427/820&$1.02^{+0.10}_{-0.10}$&5.47$\pm$0.41&$14.0^{+3.9}_{-2.0}$&$\ldots$\\
\enddata
\tablenotetext{a}{The power-law photon index and the observed gamma-ray fluence
in the BAT band (15-150keV,in units of $10^{-6}$erg cm$^{-2}$).}
\tablenotetext{b}{The time interval for our XRT light curve fitting, and the
corresponding temporal decay slope with fitting $\chi^2/$dof, time-integrated
spectral index and  hydrogen column density $N_{H}$ at the host galaxy are (in
units of $10^{20}$ $cm^{-2}$). The spectral parameters are derived from PC data
only (from Evens et al. 2008).} \tablenotetext{c}{The X-ray fluence calculated
by integrating the fitting light curve from 10 seconds after the GRB trigger to
$10^5$ s, in units of $10^{-6}$erg cm$^{-2}$.} \tablenotetext{d}{The References
of redshift.}

\tablerefs{1: Jakobsson et al.(2005); 2: Grazian et al.(2006); 3: Bloom et
al.(2006b); 4: Starling et al.(2006); 5: Osip et al.(2006); 6: Chen et
al.(2007); 7: Jakobsson et al.(2007a); 8: Fatkhullin et al.(2007); 9: Bloom et
al.(2008); 10: Cucchiara et al.(2008); 11: Thoene et al.(2008); 12: Hessels et
al.(2008); 13: de Ugarte Postigo et al.(2009)}
\end{deluxetable}


\begin{deluxetable}{lllllllllllllll}
\rotate \tablewidth{650pt}


\tabletypesize{\tiny}

\tablecaption{The observations and our fits for the GRBs with a canonical XRT
lightcurve} \tablenum{2}

\tablehead{\colhead{GRB}& \colhead{$T_{90}$(s)}&
\colhead{$\Gamma_\gamma$\tablenotemark{a}}&
\colhead{$S_\gamma$\tablenotemark{a}}& \colhead{$\alpha_1$\tablenotemark{b}}&
\colhead{$\alpha_2$\tablenotemark{b}}& \colhead{$\chi^2/$dof\tablenotemark{b}}&
\colhead{$t_b$(ks)\tablenotemark{b}}& \colhead{$S_X$\tablenotemark{b}}&
\colhead{$\beta_x$\tablenotemark{c}}& \colhead{$N_{H}$\tablenotemark{c}}&
\colhead{$T_0$(ks)\tablenotemark{d}}& \colhead{$\chi^2$/dof\tablenotemark{d}}&
\colhead{$z$}& \colhead{ref\tablenotemark{e}}}

\startdata
050319&152.5&2.02$\pm$0.19&13.10$\pm$1.48&0.47$\pm$0.36&1.11$\pm$0.15&9/11&20.57$\pm$9.32&1.34$\pm$1.54&$0.95^{+0.07}_{-0.05}$&$9.5^{+23.1}_{-9.5}$&4.93$\pm$0.45&110/72&3.24&1\\
050401&33.3&1.40$\pm$0.07&82.20$\pm$3.06&0.57$\pm$0.02&1.37$\pm$0.06&106/92&5.86$\pm$0.78&13.21$\pm$1.86&$0.99^{+0.11}_{-0.11}$&$139^{+65}_{-59}$&1.05$\pm$0.18&524/330&2.9&2\\
050416A&2.5&3.08$\pm$0.22&3.67$\pm$0.37&0.43$\pm$0.12&0.90$\pm$0.04&36/38&1.74$\pm$1.12&1.49$\pm$0.35&$1.15^{+0.07}_{-0.05}$&$54.1^{+5.8}_{-7.6}$&0.38$\pm$0.08&72/91&0.65&3\\
050505&58.9&1.41$\pm$0.12&24.90$\pm$1.79&0.15$\pm$0.19&1.30$\pm$0.06&26/45&7.87$\pm$1.57&3.18$\pm$0.56&$1.03^{+0.04}_{-0.05}$&$161.9^{+38.1}_{-21.5}$&7.15$\pm$0.41&182/191&4.27&4\\
050802&19&1.54$\pm$0.13&20.0$\pm$1.57&0.32$\pm$0.10&1.61$\pm$0.04&58/72&4.09$\pm$0.61&4.82$\pm$0.94&$0.89^{+0.03}_{-0.05}$&$2.63^{+0.99}_{-0.96}$&2.22$\pm$0.70&185/239&1.71&5\\
050824&22.6&2.76$\pm$0.38&2.66$\pm$0.52&0.32$\pm$0.06&1.00$\pm$0.05&51/39&92.22$\pm$49.75&0.93$\pm$0.14&$0.93^{+0.14}_{-0.14}$&$3.7^{+8.2}_{-3.7}$&42.41$\pm$6.32&51/41&0.83&6\\
051016B&4&2.40$\pm$0.23&1.70$\pm$0.22&0.39$\pm$0.08&1.18$\pm$0.04&15/16&18.77$\pm$3.45&2.32$\pm$1.01&$1.19^{+0.06}_{-0.13}$&$55.4^{+6.1}_{-14}$&6.34$\pm$0.65&53/61&0.94&7\\
051109A&37.2&1.51$\pm$0.20&22.0$\pm$2.72&0.79$\pm$0.07&1.53$\pm$0.08&39/48&27.28$\pm$7.90&11.32$\pm$4.75&$0.90^{+0.08}_{-0.08}$&$50.8^{+29.6}_{-27.3}$&2.43$\pm$0.27&167/167&2.35&8\\
060115&139.6&1.75$\pm$0.12&17.10$\pm$1.50&0.61$\pm$0.08&1.31$\pm$0.11&19/16&37.16$\pm$40.16&0.82$\pm$0.23&$1.06^{+0.08}_{-0.05}$&$<20.8$&11.38$\pm$2.07&20/18&3.53&9\\
060124&8.2&1.84$\pm$0.19&4.61$\pm$0.53&0.78$\pm$0.10&1.65$\pm$0.05&165/132&52.65$\pm$10.33&30.72$\pm$12.36&$1.08^{+0.05}_{-0.05}$&$65.1^{+16}_{-15.2}$&5.57$\pm$0.58&337/292&2.297&10\\
060418&103.1&1.70$\pm$0.06&83.30$\pm$2.53&0.93$\pm$0.07&1.61$\pm$0.05&83/76&1.73$\pm$1.46&2.16$\pm$1.65&$0.89^{+0.10}_{-0.09}$&$34.5^{+16.9}_{-15.6}$&0.45$\pm$0.04&84/78&1.49&11\\
060502A&28.4&1.46$\pm$0.08&23.10$\pm$1.02&0.53$\pm$0.03&1.68$\pm$0.15&11/26&72.57$\pm$15.05&5.39$\pm$1.23&$1.15^{+0.11}_{-0.10}$&$12.5^{+11.7}_{-11.1}$&6.16$\pm$0.68&64/70&1.51&12\\
060714&115&1.93$\pm$0.11&28.30$\pm$1.67&0.34$\pm$0.10&1.27$\pm$0.05&53/73&3.70$\pm$0.97&2.27$\pm$0.52&$1.04^{+0.09}_{-0.08}$&$96.3^{+32.1}_{-29.8}$&1.56$\pm$0.15&58/48&2.71&13\\
060729&115.3&1.75$\pm$0.14&26.10$\pm$2.11&0.21$\pm$0.01&1.42$\pm$0.02&459/459&72.97$\pm$3.02&20.65$\pm$0.89&$1.26^{+0.05}_{-0.05}$&$11.9^{+1.6}_{-1.5}$&51.12$\pm$11.21&1246/680&0.54&14\\
060814&145.3&1.53$\pm$0.03&146.0$\pm$2.39&0.54$\pm$0.02&1.59$\pm$0.05&81/57&17.45$\pm$1.71&8.02$\pm$0.87&$1.30^{+0.07}_{-0.07}$&$28.9^{+2.3}_{-2.2}$&5.34$\pm$0.39&228/171&0.84&15\\
060906&43.5&2.03$\pm$0.11&22.10$\pm$1.36&0.35$\pm$0.10&1.78$\pm$0.10&49/32&13.66$\pm$3.29&1.18$\pm$0.25&$1.12^{+0.13}_{-0.16}$&$245^{+138.6}_{-130.5}$&10.64$\pm$1.01&60/34&3.68&16\\
060908&19.3&1.35$\pm$0.06&28.0$\pm$1.11&0.70$\pm$0.10&1.49$\pm$0.08&36/30&0.95$\pm$0.34&2.25$\pm$1.20&$1.00^{+0.09}_{-0.08}$&$26.2^{+22.2}_{-20}$&0.35$\pm$0.03&46/32&2.43&17\\
060912&5&1.74$\pm$0.09&13.50$\pm$0.62&0.13$\pm$0.30&1.19$\pm$0.08&22/28&1.13$\pm$0.31&0.92$\pm$0.60&$0.95^{+0.11}_{-0.06}$&$24.2^{+8.2}_{-5}$&0.21$\pm$0.06&25/30&0.94&18\\
060927&22.5&1.65$\pm$0.08&11.30$\pm$0.68&0.60$\pm$0.05&1.76$\pm$0.20&8/13&3.04$\pm$1.41&0.73$\pm$0.21&$0.89^{+0.17}_{-0.16}$&$117.7^{+197.5}_{-117.7}$&1.22$\pm$0.12&14/16&5.6&19\\
061121&81.3&1.41$\pm$0.03&137.0$\pm$1.99&0.75$\pm$0.06&1.63$\pm$0.05&121/147&24.32$\pm$4.38&20.35$\pm$5.65&$0.93^{+0.04}_{-0.04}$&$49.2^{+5.5}_{-5.2}$&3.09$\pm$0.68&281/285&1.31&20\\
070306&209.5&1.66$\pm$0.10&53.80$\pm$2.86&0.12$\pm$0.02&1.87$\pm$0.03&102/114&29.69$\pm$1.72&7.84$\pm$0.42&$1.12^{+0.08}_{-0.08}$&$33.4^{+3.1}_{-3}$&29.45$\pm$1.26&140/233&1.497&21\\
070508&20.9&1.35$\pm$0.03&196.0$\pm$2.73&0.45$\pm$0.02&1.42$\pm$0.01&516/489&0.77$\pm$0.22&19.50$\pm$1.53&$0.75^{+0.09}_{-0.09}$&$23.4^{+3.8}_{-3.6}$&0.43$\pm$0.02&511/491&0.82&22\\
070529&109.2&1.34$\pm$0.16&25.70$\pm$2.45&0.75$\pm$0.06&1.32$\pm$0.04&30/32&2.25$\pm$1.55&1.74$\pm$0.31&$0.89^{+0.09}_{-0.10}$&$157.9^{+48.4}_{-62.4}$&0.44$\pm$0.06&31/34&2.4996&23\\
070810A&11&2.04$\pm$0.14&6.90$\pm$0.60&0.51$\pm$0.07&1.31$\pm$0.06&24/31&1.72$\pm$0.72&1.20$\pm$0.26&$1.06^{+0.11}_{-0.10}$&$51.5^{+15.8}_{-19.9}$&0.61$\pm$0.08&24/34&2.17&24\\
071003&150&1.36$\pm$0.07&83.0$\pm$3.0&-0.20$\pm$0.33&1.84$\pm$0.04&76/66&30.27$\pm$5.16&2.54$\pm$0.65&$1.04^{+0.13}_{-0.15}$&$4.8^{+3.2}_{-2.8}$&7.77$\pm$1.75&93/68&1.1&25\\
080210&45&1.77$\pm$0.12&18.0$\pm$1.0&0.84$\pm$0.05&1.47$\pm$0.06&45/26&7.04$\pm$5.08&2.10$\pm$1.06&$1.22^{+0.12}_{-0.10}$&$149^{+72}_{-67}$&1.02$\pm$0.14&35/27&2.64&26\\
080310&365&2.32$\pm$0.16&23.0$\pm$2.0&0.27$\pm$0.06&1.58$\pm$0.04&68/65&10.93$\pm$1.17&2.07$\pm$0.25&$1.45^{+0.02}_{-0.02}$&$70^{+10}_{-10}$&7.06$\pm$0.62&75/67&2.4266&27\\
080430&16.2&1.73$\pm$0.09&12.0$\pm$1.0&0.46$\pm$0.02&1.17$\pm$0.02&81/140&33.20$\pm$6.32&2.73$\pm$0.14&$1.05^{+0.07}_{-0.07}$&$33.5^{+4.3}_{-4}$&8.66$\pm$0.57&108/142&0.767&28\\
080516&5.8&1.82$\pm$0.27&2.60$\pm$0.40&0.29$\pm$0.07&1.00$\pm$0.06&24/26&2.30$\pm$1.17&0.88$\pm$0.18&$1.26^{+0.18}_{-0.24}$&$58.3^{+21.8}_{-15.9}$&1.12$\pm$0.17&24/29&3.2&29\\
080605&20&1.11$\pm$0.14&133.0$\pm$2.0&0.58$\pm$0.04&1.43$\pm$0.02&359/309&0.60$\pm$0.07&15.23$\pm$2.52&$0.76^{+0.09}_{-0.08}$&$65.7^{+19.7}_{-18.2}$&0.26$\pm$0.01&345/311&1.6398&30\\
080707&27.1&1.77$\pm$0.19&5.20$\pm$0.60&0.22$\pm$0.05&1.07$\pm$0.05&21/21&7.80$\pm$3.48&0.74$\pm$0.25&$1.11^{+0.17}_{-0.18}$&$38^{+19}_{-19}$&4.37$\pm$0.74&25/22&1.23&31\\
080710&120&1.47$\pm$0.23&14.0$\pm$2.0&0.94$\pm$0.05&1.80$\pm$0.17&79/63&21.30$\pm$5.55&4.32$\pm$1.03&$1.05^{+0.07}_{-0.09}$&$12.5^{+4.9}_{-4.5}$&8.17$\pm$0.65&81/64&0.845&32\\
080721&16.2&1.11$\pm$0.08&120.0$\pm$10.0&0.80$\pm$0.01&1.65$\pm$0.01&1534/1371&3.07$\pm$0.34&69.07$\pm$4.66&$0.81^{+0.03}_{-0.03}$&$62.6^{+13}_{-12.5}$&0.57$\pm$0.07&1899/1373&2602&33\\
080905B&128&1.78$\pm$0.15&18.0$\pm$2.0&0.11$\pm$0.04&1.42$\pm$0.02&81/74&2.93$\pm$0.72&8.87$\pm$0.87&$0.94^{+0.10}_{-0.06}$&$238^{+51}_{-43}$&2.28$\pm$0.13&104/74&2.347&34\\
081007&10&2.51$\pm$0.20&7.10$\pm$0.80&0.72$\pm$0.02&1.35$\pm$0.05&68/61&56.35$\pm$21.72&2.41$\pm$0.85&$1.14^{+0.12}_{-0.12}$&$50.9^{+8.5}_{-7.3}$&4.19$\pm$0.26&105/92&0.5295&35\\
081008&185.5&1.69$\pm$0.07&43.0$\pm$2.0&0.82$\pm$0.03&1.86$\pm$0.08&30/44&15.71$\pm$3.59&3.56$\pm$0.53&$1.06^{+0.11}_{-0.06}$&$24^{+21}_{-12}$&4.01$\pm$0.27&41/46&1.967&36\\
081203A&294&1.54$\pm$0.06&77.0$\pm$3.0&1.13$\pm$0.20&2.06$\pm$0.33&232/221&11.49$\pm$1.97&13.50$\pm$3.01&$1.09^{+0.09}_{-0.08}$&$54^{+17}_{-15}$&0.72$\pm$0.34&304/222&2.1&37\\
081222&24&1.08$\pm$0.15&48.0$\pm$1.0&0.63$\pm$0.07&1.15$\pm$0.02&459/385&0.31$\pm$0.11&8.59$\pm$1.12&$1.06^{+0.07}_{-0.07}$&$60^{+20}_{-19}$&0.17$\pm$0.01&469/386&2.77&38\\
090418A&56&1.48$\pm$0.07&46.0$\pm$2.0&0.38$\pm$0.02&1.61$\pm$0.03&84/108&2.68$\pm$0.32&6.67$\pm$0.94&$1.09^{+0.09}_{-0.09}$&$115^{+19}_{-18}$&1.48$\pm$0.16&96/109&1.608&39\\
090423&10.3&0.80$\pm$0.50&5.90$\pm$0.40&-0.16$\pm$0.07&1.43$\pm$0.04&25/39&4.35$\pm$0.71&1.19$\pm$0.16&$0.83^{+0.10}_{-0.10}$&$640^{+280}_{-210}$&4.12$\pm$0.54&64/40&8.1&40\\
090424&48&1.19$\pm$0.15&210.0$\pm$0.0&0.53$\pm$0.05&1.20$\pm$0.10&586/506&1.06$\pm$0.25&32.91$\pm$2.22&$0.99^{+0.08}_{-0.08}$&$45^{+5.8}_{-5.4}$&0.65$\pm$0.32&591/507&0.544&41\\
090516A&210&1.84$\pm$0.11&90.0$\pm$6.0&0.75$\pm$0.05&1.84$\pm$0.04&139/133&16.32$\pm$2.73&4.62$\pm$1.01&$1.15^{+0.05}_{-0.06}$&$203^{+38}_{-30}$&1.09$\pm$0.26&158/134&4.109&42\\
090529&$>$100&2.01$\pm$0.30&6.80$\pm$1.70&-0.18$\pm$0.23&1.13$\pm$0.11&3/5&31.33$\pm$15.04&0.51$\pm$0.33&$1.16^{+0.12}_{-0.09}$&$43^{+22}_{-21}$&32.16$\pm$15.86&4/6&2.625&3\\
090618&113.2&1.42$\pm$0.08&1050$\pm$10&0.71$\pm$0.01&1.49$\pm$0.01&1130/1078&7.89$\pm$0.45&41.71$\pm$3.03&$1.01^{+0.05}_{-0.05}$&$22.4^{+2.6}_{-2.5}$&1.63$\pm$0.45&1450/1079&0.54&44\\
\enddata
\tablenotetext{a}{TThe power-law photon index and the observed gamma-ray
fluence in the BAT band (15-150keV,in units of $10^{-7}$erg cm$^{-2}$).}

\tablenotetext{b}{The decay slopes and the break time of the shallow-to-normal
transition in the XRT lightcurve derived from  a smooth broken power law fit
with fitting $\chi^2/$dof. The $X$-ray fluences in the XRT band (0.3-10 keV)
are integrated from 10 seconds post the GRB trigger to $10^5$ seconds,in units
of $10^{-7}$erg cm$^{-2}$.}

\tablenotetext{c}{The spectral parameters of the absorbed power law model(from
Evens et al. 2008). The $N_{H}$ of the host galaxy are in units of $10^{20}$
$cm^{-2}$. }

\tablenotetext{d}{The zero time $|T_0|$ of the prior X-rays with respect to the
GRB trigger time derived from a power-law fit (Eq. 2 in the text).}

\tablenotetext{f}{The reference of redshift.}

\tablerefs{1: Fynbo et al.(GCN 3136); 2: Fynbo et al.(GCN 3176); 3: Cenko et
al.(GCN 3542); 4: Berger et al.(GCN 3368); 5: Fynbo et al.(GCN 3749); 6: Fynbo
et al.(GCN 3874); 7: Soderberg et al.(GCN 4186); 8: Quimby et al.(GCN 4221); 9:
Piranomonte et al.(GCN 4520); 10: Cenko et al.(GCN 4592); 11: Dupree et al.(GCN
4969); 12: Cucchiara et al.(GCN 5052); 13: Jakobsson et al.(GCN 5320); 14:
Thoene et al.(GCN 5373); 15: Thoene et al.(GCN 6663); 16: Vreeswijk et al.(GCN
5535); 17: Rol et al.(GCN 5555); 18: Jakobsson et al.(GCN 5617); 19: Fynbo  et
al.(GCN 5651); 20: Bloom et al.(GCN 5826); 21: Jaunsen  et al.(GCN 6202); 22:
Jakobsson  et al.(GCN 6398); 23: Chandra  et al.(GCN 6740); 24: Thoene  et
al.(GCN 6741); 25: Perley et al.(GCN 6850); 26:  Cucchiara et al.(GCN 7290);
27: Prochaska et al.(GCN 7388); 28: Cucchiara et al.(GCN 7654); 29: Filgas  et
al.(GCN 7747); 30: Jakobsson  et al.(GCN 7832); 31: Fynbo et al.(GCN 7949); 32:
Perley et al.(GCN 7962); 33: D'Avanzo et al.(GCN 7997); 34: Vreeswijk et
al.(GCN 8191); 35: Berger et al.(GCN 8335); 36: Cucchiara et al.(GCN 8346); 37:
Landsman et al.(GCN 8601); 38: Cucchiara et al.(GCN 8713); 39: Chornock et
al.(GCN 9151); 40: Fernandez-Soto et al.(GCN 9222); 41: Chornock et al.(GCN
9243); 42: de Ugarte Postigo et al.(GCN 9383); 43: Malesani et al.(GCN 9457);
44: Cenko et al.(GCN 9518);}
\end{deluxetable}


\clearpage \thispagestyle{empty} \setlength{\voffset}{-18mm}

\clearpage \setlength{\voffset}{0mm}

\begin{figure*}
\includegraphics[angle=0,scale=0.50]{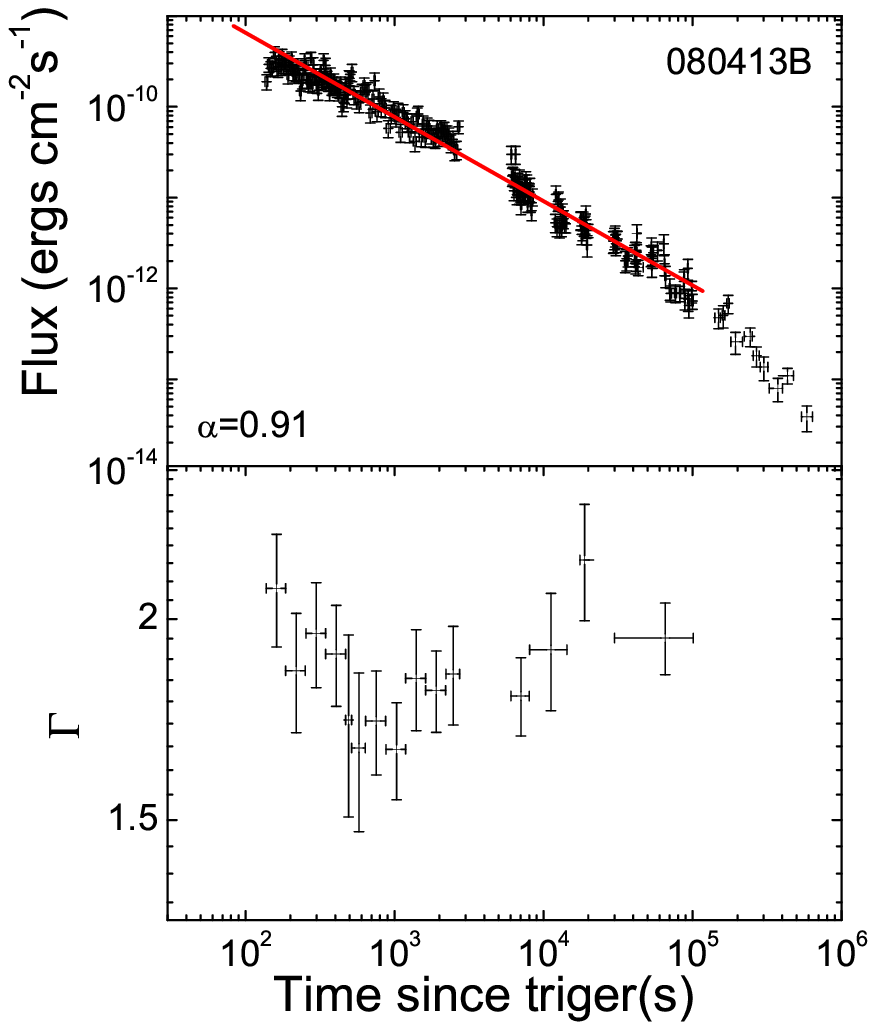}
\includegraphics[angle=0,scale=0.50]{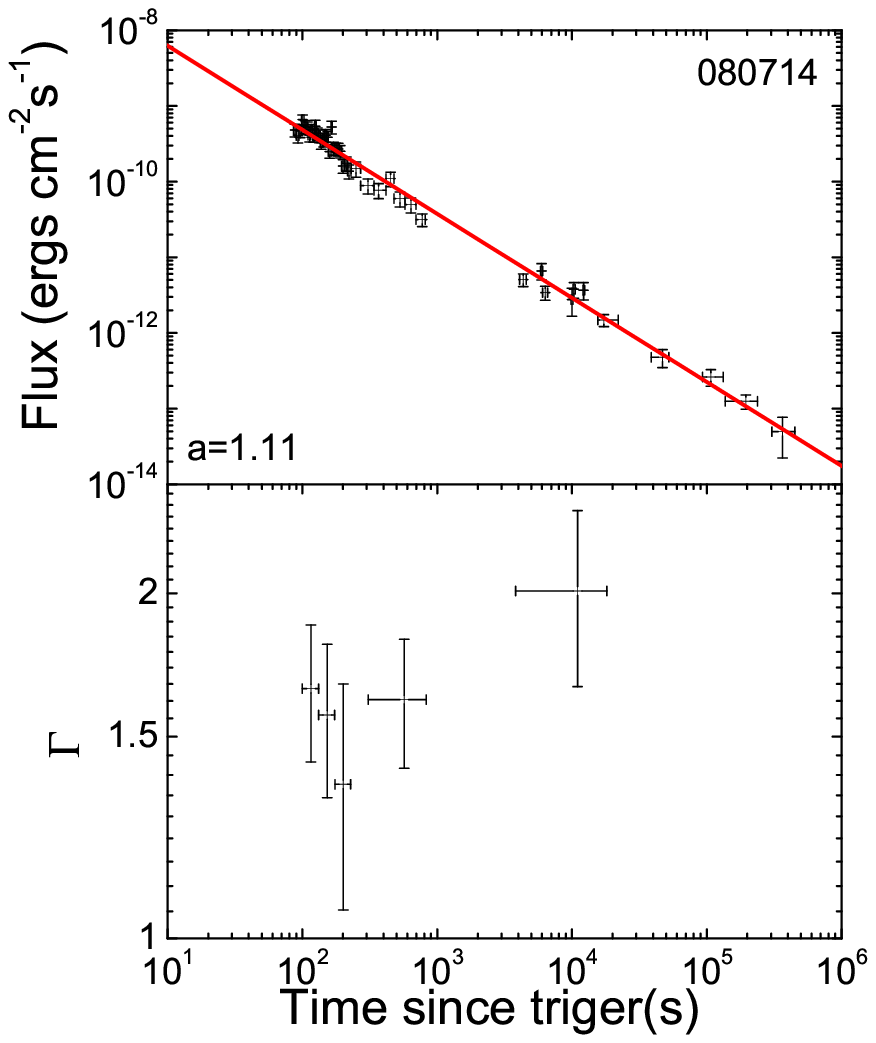}
\includegraphics[angle=0,scale=0.50]{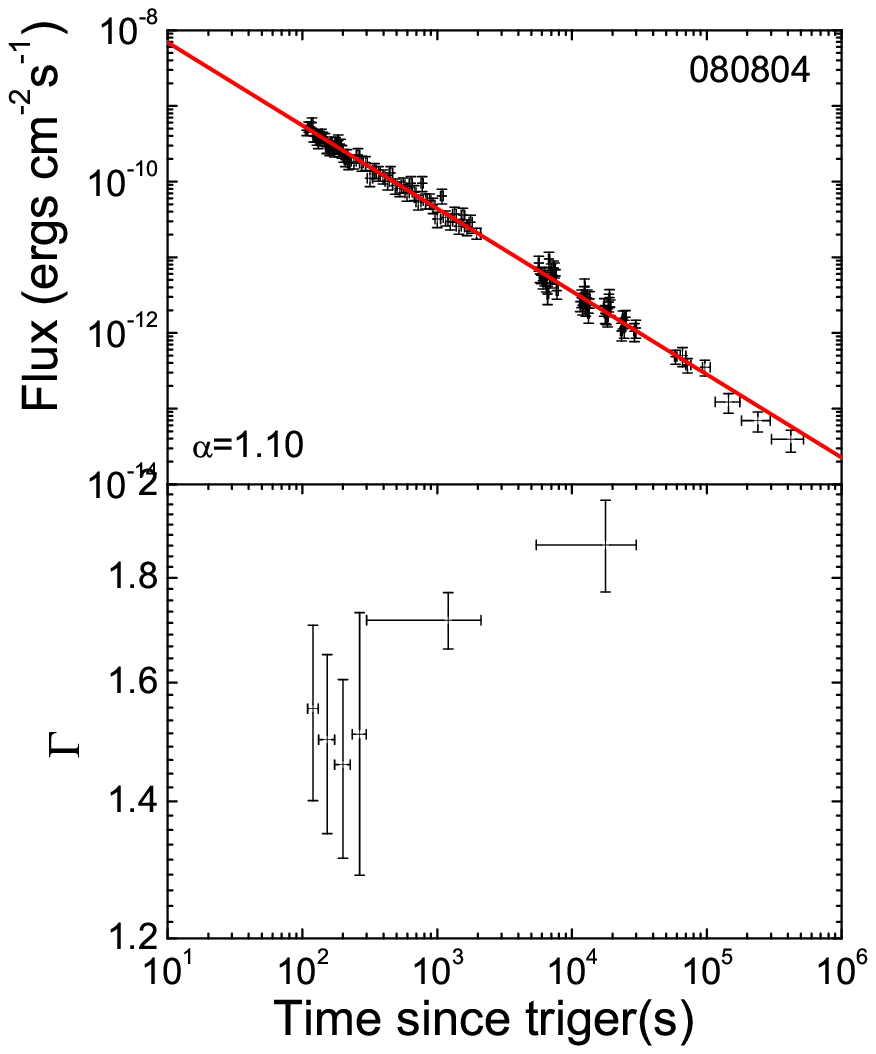}
\includegraphics[angle=0,scale=0.50]{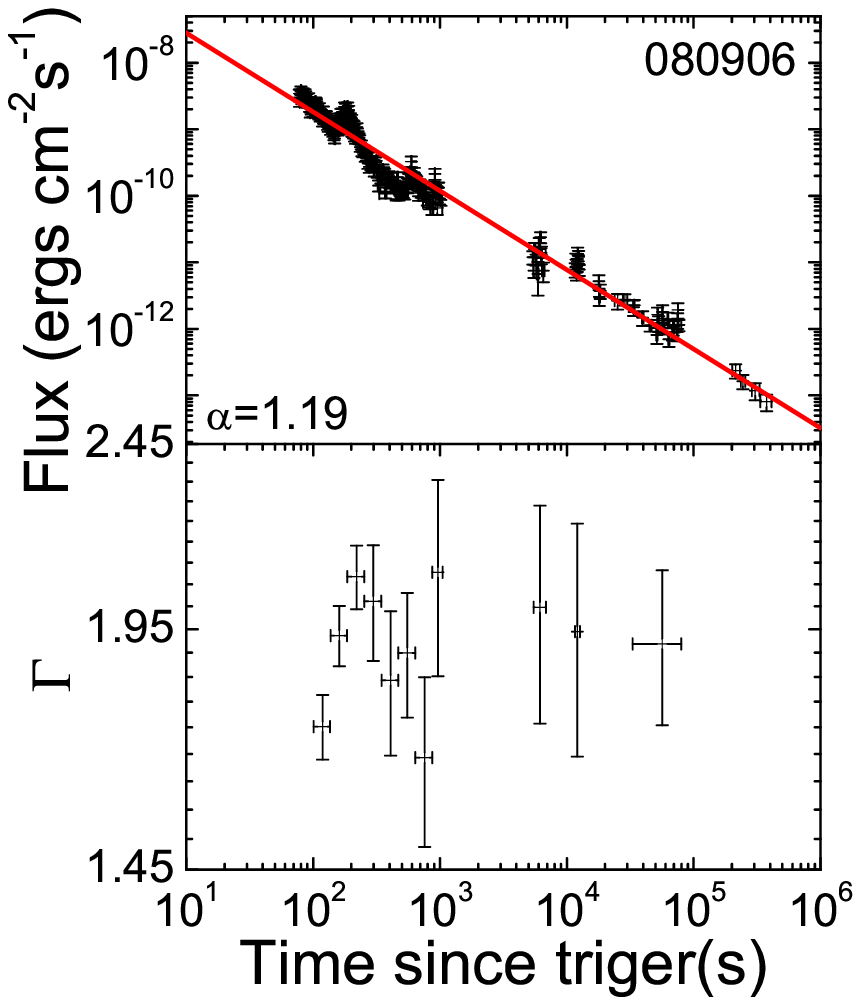}
\includegraphics[angle=0,scale=0.50]{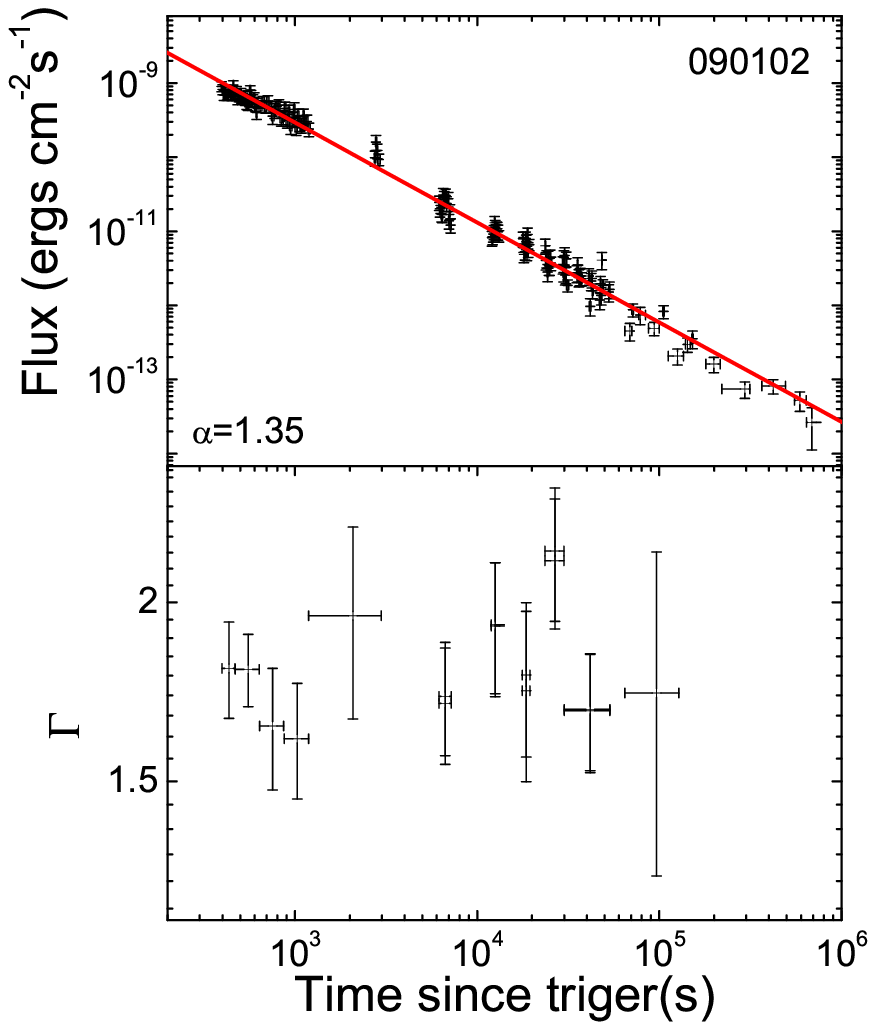}
\includegraphics[angle=0,scale=0.50]{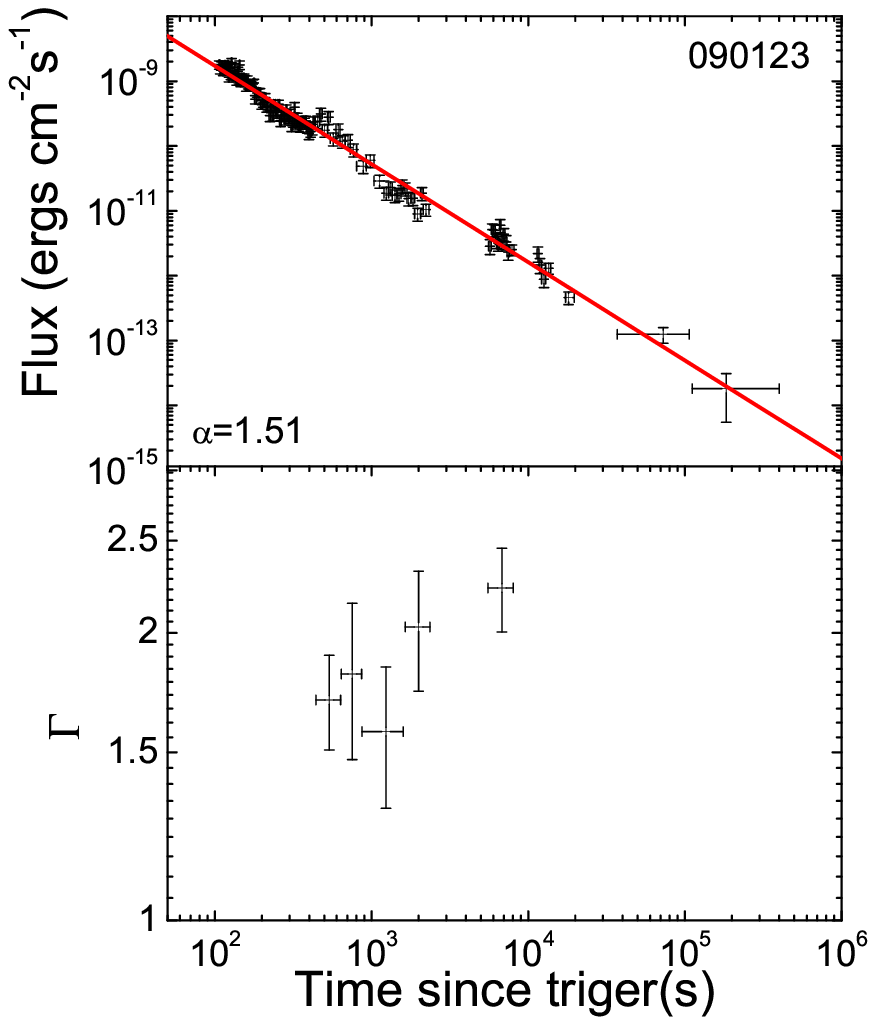}
\includegraphics[angle=0,scale=0.50]{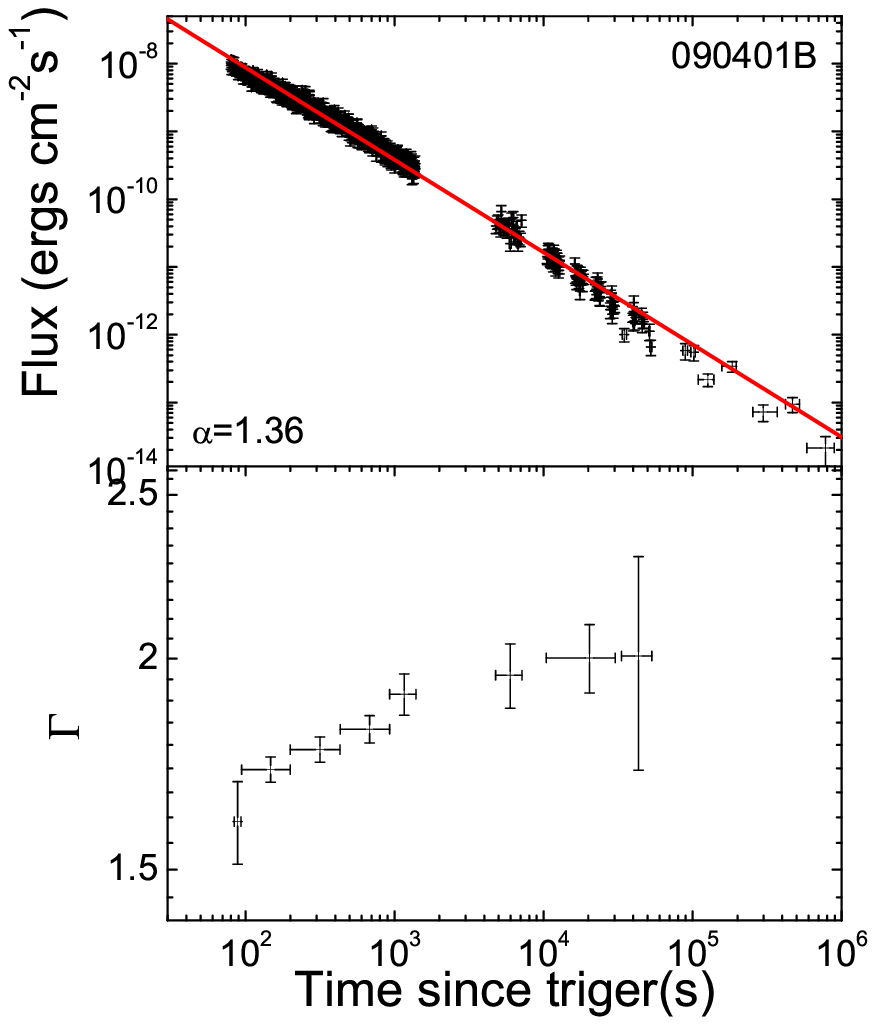}
\hfill\center{Fig. 1---  continued}
\end{figure*}


\clearpage \thispagestyle{empty} \setlength{\voffset}{-18mm}

\clearpage

\clearpage

\clearpage

\clearpage

\clearpage

\clearpage

\clearpage

\end{document}